\documentclass[10pt, aps, showpacs, preprintnumbers, prd, nofootinbib, preprint, twocolumn]{revtex4}
\usepackage{feynmf}
\usepackage{amsmath}
\usepackage{graphicx}
\usepackage{mathrsfs}
\usepackage{stmaryrd}
\usepackage{amsfonts}
\usepackage{wrapfig}
\usepackage{subfig}
\usepackage{float}
\usepackage{bbm}
\usepackage{bbold}
\usepackage{hyperref}
\usepackage{cleveref}

\begin{document}
\def\sech{\mathop{\rm{sech}}\nolimits}
\def\arcsinh{\mathop{\rm{arcsinh}}\nolimits}

\title{Solutions for Intersecting Domain Walls with Internal Structure in Six Dimensions from a $\mathbb{Z}_2\times{}\mathbb{Z}_2$-invariant Action}
\author{Benjamin D. Callen}\email{bdcallen@student.unimelb.edu.au}
\affiliation{ARC Centre of Excellence for Particle Physics at the Terascale, School of Physics, The University of Melbourne, Victoria 3010, Australia}
\author{Raymond R. Volkas}\email{raymondv@unimelb.edu.au}
\affiliation{ARC Centre of Excellence for Particle Physics at the Terascale, School of Physics, The University of Melbourne, Victoria 3010, Australia}

\begin{abstract}

 We present a generic $\mathbb{Z}_{2}\times{}\mathbb{Z}_{2}$-invariant scalar field theory with four real scalar fields in six-dimensional Minkowskian spacetime which yields solutions consisting of two intersecting domain-wall kinks which are each paired by fields with lump-like profiles. For a special parameter choice, analytic solutions can be obtained. We show that the $\mathbb{Z}_{2}\times{}\mathbb{Z}_{2}$ symmetry can be maintained while coupling fermions by introducing scalar Yukawa couplings to one kink-lump pair and six-dimensional pseudoscalar Yukawa couplings to the other, and we show that there exists a zero mode localized to the domain-wall junction in this case. We also show that scalar fields can be localized.

\end{abstract}

\maketitle

\newpage

\section{Introduction}
\label{sec:introduction}

 The possibility that there could be hidden extra dimensions of space continues to be an intriguing one for describing physics beyond the Standard Model (SM). Most extra-dimensional models, such as string theory, use compact and tiny Planck-scale extra dimensions. Recently, large extra dimensions of order an inverse TeV have been considered \cite{originalbranepaper, antoniadisnewdimattev, newdimatmillimeter, exotickkmodels, gibbonswiltshire, pregeometryakama}, which became particularly popular with the advent of the model by Arkani-Hamad, Dimopolous and Dvali (ADD) which featured a solution to the hierarchy problem. Later, Randall and Sundrum proposed the RS1 model \cite{randallsundrum2} which presents an alternative solution to the hierarchy problem in a slice of warped anti-de Sitter (AdS) geometry. The same two authors showed that infinite extra dimensions were possible, in a model now termed RS2 \cite{randallsundrum1}. In this model, matter is confined to a fundamental positive-tension brane put into the theory by hand, and gravity is localized onto this fundamental brane. 

 One could consider the possibility that such a brane on which matter and gravity is localized is not put in by hand but rather generated dynamically by the underlying theory. Such an object must be a stable solution to the theory; thus the obvious candidates for such an object are solitons. Topological solitons are classical solutions to the dynamical equations of a theory which non-trivially map a boundary of one or more dimensions of spacetime at infinity to the moduli space of vacua. These mappings belong to the non-trivial homotopy classes of mappings between these two spaces. Such topological solitons are stable against decay to a trivial vacuum due to the existence of conserved topological charges which are associated with these homotopy classes. Examples of solitons are domain walls which are classical solutions which map spatial infinity along both directions of a particular dimension to two distinct but degenerate vacua, strings which are non-trivial maps from the boundary of a two-dimensional 
space to a moduli space with the geometry of a circle, and monopoles which are non-trivial mappings from a sphere at spatial infinity to a sphere of vacua. Localizing matter on solitons, particularly domain walls, is not a new idea and the proposition of the world volume of a domain wall forming our observable 3+1-dimensional (3+1D) universe was first proposed by Rubakov and Shaposhnikov \cite{rubshapdwbranes}.

 In addition to the original proposal by Rubakov and Shaposhnikov, there have been several other attempts at demostrating the viability of such a model based on domain-wall branes. Localization of fermions and scalars can be easily achieved by introducing appropriate Yukawa and quartic scalar interactions respectively (for an analysis of the localization of scalar and fermionic zero modes as well as their Kaluza-Klein (KK) spectra, see \cite{modetower}). For the localization of gauge bosons, Dvali and Shifman have proposed a mechanism by which a gauge group $G$ is broken down to a subgroup $H$ in the interior of the domain wall and is unbroken and confining in the bulk, localizing the gauge bosons of $H$ to the domain wall through confinement dynamics \cite{dsmech}. Importantly, gravity can be localized on the domain wall, yielding an RS2-like warped background geometry and in its presence, scalars and fermions can still be localized \cite{rsgravitydaviesgeorge2007}. A viable model utilizing the Dvali-
Shifman mechanism with $G=SU(5)$ and $H=SU(3)\times{}SU(2)\times{}U(1)$ in 4+1D spacetime was proposed in \cite{firstpaper} which has some very interesting phenomenology; the SM components embedded within the $SU(5)$ multiplets are split and localized about different points in the extra dimension leading to a natural realization of the split fermion mechanism first proposed by Arkani-Hamed and Schmaltz \cite{splitfermions}. Furthermore, such splitting of fermions and scalars within the model can account for the fermion mass hierarchy, quark mixing and the suppression of proton decay \cite{su5branemassfittingpaper} and a simple extension of the model to include the discrete flavor group $A_{4}$ can naturally generate large lepton mixing angles \cite{su5a4braneworldcallen}. Models based on extended gauge groups such as $SO(10)$ \cite{jayneso10paper} and $E_{6}$ \cite{e6domainwallpaper} are also possible. In this paper, we propose an extension involving the addition of a second extra dimension with matter being 
localized to the intersection of two domain walls. 

 Topological defects of co-dimension 2 have been considered as candidates for the localization of fields onto a 3+1D subspace. The simplest example of a co-dimension 2 defect is a string and it has already been shown that 3+1D gravity can be reproduced on such an object in a 5+1D spacetime \cite{warpedgravityonvortex, gherghetta6dstring}. Extensions of the RS2 model for the intersection of $n$ fundamental branes in 4+n-dimensional spacetime were proposed in \cite{arkanihamedinfinitenewdimensions, flachimasato6dintbrane}. Other than strings, we could instead consider the possibility of using a pair of domain walls in 6D spacetime to localize fields on to a 4D world volume. 

 There are two ways of introducing a second domain wall in order to freeze out a second extra dimension. One way is to localize a scalar field onto a domain wall and have that scalar field develop a tachyonic mass which induces a breaking of a second discrete symmetry, yielding a second domain wall localized to the first. This is called a nested domain wall or domain ribbon and there is some literature that has dealt with this scenario \cite{morrisnestedwalls, britobazeiadomainribbon} and with the possible localization of gravity on such a defect \cite{nestedbranegravity}. A second way to produce a dimensional reduction from 5+1D to 3+1D is to have two stable domain walls which intersect. There have been some supersymmetric models which yield rare exact solutions for a pair of non-trivially intersecting walls \cite{hksigmamodelintersectingwalls}. For more on models with intersecting domain walls or domain-wall junctions, see \cite{exactdomainwalljunction, dwjunctionnonnormmodes, bpseqintersectingwall, troitskyvoloshinintersectingdw}.

 In this paper, we present a $\mathbb{Z}_{2}\times{}\mathbb{Z}_{2}$-invariant 5+1D interacting scalar field theory with four real scalar fields, where the 6D masses are tachyonic for two of the scalar fields and positive definite for the other two, in which a rare analytic solution for two intersecting domain walls can be obtained for a particular parameter choice. We find that there exists a class of energy degenerate solutions with two domain walls: one in which the walls are perpendicular, a range of solutions where the walls intersect at an angle between 0 and 90 degrees and another in which the walls are parallel. We give topological arguments for why the perpendicular solution cannot evolve to the parallel solution given for the particular parameter choice, as well as an argument for why the perpendicular solution might be energetically favorable to the solutions with intersection angle less than 90 degrees in a nearby region of parameter space (assuming they are not topologically distinct). We also show that chiral fermions and scalars can be localized to the intersection of the domain wall. This is important because these chiral fermions and scalars form the building blocks of the quarks, leptons and Higgs bosons of an effective Standard Model-like theory dynamically localized to the intersection.

 In Sec. \ref{sec:kinklumpreview}, we give a review of domain walls with an additional scalar field which attains a tachyonic mass in the interior of the domain wall and which thus condenses to attain a lump-like vacuum expectation value profile. In Sec. \ref{sec:model}, we outline the model generating the intersecting domain wall solution, give the form of the solution for which the walls are perpendicular and we also give a topological argument for why this solution cannot evolve to the solution where the walls are parallel despite these solutions being energy degenerate. We also give an argument for why the perpendicular solution and the solutions which intersect at an angle between 0 and 90 degrees are not energy degenerate in general. In Sec. \ref{sec:fermionlocalization}, we discuss fermion localization and we show that localization of a single chiral zero mode on the intersection of the domain walls is possible. In Sec. \ref{sec:scalarlocalization}, we show that scalars can also be localized to the domain-wall intersection. Section \ref{sec:conclusion} is our conclusion.

\section{Domain Walls with Internal Structure: A Review}
\label{sec:kinklumpreview}

 Before we do the full analysis for the intersecting domain-wall solution, we present a brief overview of domain-wall solutions with internal structure. Domain walls with internal structure are common in interacting scalar field theories involving at least two real scalar fields. They consist of one scalar field which generates the domain wall and thus has a kink-like profile, and one or more scalar fields which interact with the kink field attaining a lump-like profile. One can think of these extra fields as being dynamically localized to the wall which at the same time attain tachyonic masses in the interior of the wall and thus non-zero vacuum expectation values. In general, if the back reactions of these fields on the kink field are significant, these can affect the domain-wall solution, particularly its width.

 To present an example of a domain wall with internal structure, we write down a $\mathbb{Z}_{2}$-symmetric scalar field theory with two scalar fields, $\eta$ and $\chi$. The field $\eta$ will form the background domain wall, and $\chi$ will condense in the interior. The role of the $\mathbb{Z}_2$ symmetry is to ensure topological stability after its spontaneous breaking. We desire to find a solution such that $\eta$ interpolates between two values $\pm{}v$ from $y=-\infty$ to $y=+\infty$ and where the vacuum expection value of $\chi$ tends to zero out at infinity. To generate such a solution and ensure its topological stability, we must ensure that $\eta=\pm{}v$, $\chi=0$ are global minima of the potential. Hence, we give $\eta$ a tachyonic mass, while the mass squared of $\chi$ will be made positive definite. Given these requirements and the $\mathbb{Z}_2$ symmetry $\eta\rightarrow-\eta$, $\chi\rightarrow-\chi$, the scalar potential is

\begin{equation}
\begin{aligned}
\label{eq:singlekinklumpscalarpotential}
V(\eta, \chi) &= \frac{1}{4}\lambda_{\eta}(\eta^2-v^2)^2+\frac{1}{2}\lambda_{\eta\chi}(\eta^2-v^2)\chi^2+\frac{1}{2}\mu^2_{\chi}\chi^2 \\
              &+\frac{1}{4}\lambda_{\chi}\chi^4+g_{\eta\chi}\eta^3\chi+h_{\eta\chi}\eta\chi^3.
\end{aligned}
\end{equation}

 To ensure that the potential is bounded from below as well as to make sure we have the desired properties for $\eta$ and $\chi$, we make the following parameter choices,
\begin{equation}
\label{eq:boundednessconditions}
\lambda_{\eta}>0, \qquad{} \lambda_{\chi}>0, \qquad \lambda_{\eta\chi}v^2>\mu^2_{\chi}.
\end{equation}
The last of the above conditions ensures that the squared mass of the field $\chi$ becomes tachyonic in the interior of the domain wall, near where $\eta=0$, so that the solution where $\chi$ forms a lump is the stable solution.

\begin{figure}[h]
\includegraphics[scale=1.0]{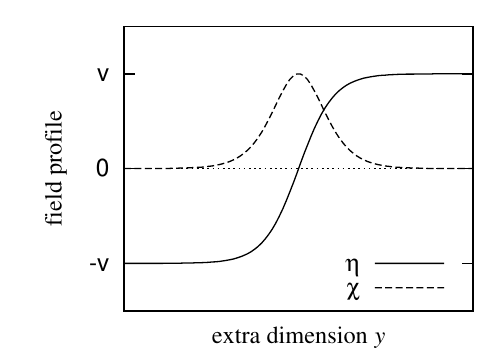}
\caption{A plot of the profiles for $\eta$ and $\chi$.}
\label{fig:singlekinklumpplot}
\end{figure}

 To form the kink-lump background, we need to look for a static solution to the Euler-Lagrange equations
\begin{equation}
\begin{aligned}
\label{eq:singlekinklumpkgequation}
\Box{}\eta&+\lambda_{\eta}(\eta^2-v^2)\eta+\lambda_{\eta\chi}\eta\chi^2+3g_{\eta\chi}\eta^2\chi+h_{\eta\chi}\chi^3=0, \\
\Box{}\chi&+\lambda_{\chi}\chi^3+\mu^2_{\chi}\chi+\lambda_{\eta\chi}(\eta^2-v^2)\chi+g_{\eta\chi}\eta^3+3h_{\eta\chi}\eta^2\chi=0,
\end{aligned}
\end{equation}
subject to the boundary conditions
\begin{equation}
\begin{aligned}
\label{eq:singlekinklumpbc}
\eta{}(y=\pm{}\infty)&=\pm{}v, \\
\chi{}(y=\pm{}\infty)&=0,
\end{aligned}
\end{equation}
that depends solely on the coordinate $y$ so that $\eta=\eta(y)$ and $\chi=\chi(y)$. This equation can be solved numerically and in general the profile for $\eta$ is kink-like and the profile for $\chi$ is indeed lump-like. For the special parameter choice 
\begin{equation}
\label{eq:singlekinklumpconditions}
g_{\eta\chi}=h_{\eta\chi}=0, \\
2\mu^2_{\chi}(\lambda_{\eta\chi}-\lambda_{\chi})+(\lambda_{\eta}\lambda_{\chi}-\lambda^2_{\eta\chi})v^2=0,
\end{equation}
one finds the analytic solution
\begin{equation}
\begin{aligned}
\label{eq:singlekinklumpsolution}
\eta(y)&=v\tanh{(ky)}, \\
\chi(y)&=A\sech{(ky)},
\end{aligned}
\end{equation}
where $k^2=\mu^2_{\chi}$ and $A^2=\frac{\lambda_{\eta\chi}v^2-2\mu^2_{\chi}}{\lambda_{\chi}}$. A plot of the solution is given in Fig.\ \ref{fig:singlekinklumpplot}.

 The solution of Eq. \ref{eq:singlekinklumpsolution} is well known to be globally and locally stable \cite{dgeorgesurvivalzeromodes, jaynestabilitypaper}.

\section{The Background Scalar Field Theory}
\label{sec:model}

 Our model describing the background fields is a 5+1-dimensional (5+1D) quartic scalar field theory with four scalar fields invariant under a $\mathbb{Z}_2\times{}\mathbb{Z}_2$ symmetry. The role of the $\mathbb{Z}_2\times{}\mathbb{Z}_2$ symmetry is to ensure topological stability of the resultant intersecting walls in a manner analogous to that of the $\mathbb{Z}_{2}$ symmetry in the single wall case. Under this symmetry,  we assign the following parities to these fields:
\begin{equation}
\begin{aligned}
\label{eq:z2xz2parities}
\eta_1 &\sim (-, +) \qquad{} \chi_1 \sim (-, +), \\
\eta_2 &\sim (+, -) \qquad{} \chi_2 \sim (+, -).
\end{aligned}
\end{equation}
 The fields $\eta_1$ and $\eta_2$ will form the two perpendicular background domain-wall kinks while $\chi_{1}$ and $\chi_{2}$ will attain lump-like profiles parallel to each respective wall. 

  Given the parity assignments, we may write the most general quartic scalar potential of this theory as 

\begin{equation}
\begin{aligned}
\label{eq:scalarpotential}
V_{DW} &= \frac{1}{4}\lambda_{\eta_1}(\eta^2_1-v^2_1)^2+\frac{1}{2}\lambda_{\eta_1\chi_1}(\eta^2_1-v^2_1)\chi^2_1+\frac{1}{2}\mu^2_{\chi_1}\chi^2_1 \\
       &+\frac{1}{4}\lambda_{\chi_1}\chi^4_{1}+g_{\eta_1\chi_1}\eta^3_1\chi_1+h_{\eta_1\chi_1}\eta_1\chi^3_1 \\
       &+ \frac{1}{4}\lambda_{\eta_2}(\eta^2_2-v^2_2)^2+\frac{1}{2}\lambda_{\eta_2\chi_2}(\eta^2_2-v^2_2)\chi^2_2+\frac{1}{2}\mu^2_{\chi_2}\chi^2_2 \\
       &+\frac{1}{4}\lambda_{\chi_2}\chi^4_{2}+g_{\eta_2\chi_2}\eta^3_2\chi_2+h_{\eta_2\chi_2}\eta_2\chi^3_2 \\
       &+\frac{1}{2}\lambda_{\eta_1\eta_2}(\eta^2_1-v^2_1)(\eta^2_2-v^2_2)+\frac{1}{2}\lambda_{\eta_1\chi_2}(\eta^2_1-v^2_1)\chi^2_2 \\
       &+\frac{1}{2}\lambda_{\chi_1\eta_2}\chi^2_1(\eta^2_2-v^2_2)+\frac{1}{2}\lambda_{\chi_1\chi_2}\chi^2_1\chi^2_2  \\
       &+\frac{1}{2}\lambda_{\eta_1\eta_2\chi_2}\eta^2_1\eta_2\chi_2+\frac{1}{2}\lambda_{\chi_1\eta_2\chi_2}\chi^2_1\eta_2\chi_2 \\
       &+\frac{1}{2}\lambda_{\eta_1\chi_1\eta_2}\eta_1\chi_1\eta^2_2+\frac{1}{2}\lambda_{\eta_1\chi_1\chi_2}\eta_1\chi_1\chi^2_2 \\
       &+\lambda_{\eta_1\chi_1\eta_2\chi_2}\eta_1\chi_1\eta_2\chi_2.
\end{aligned}
\end{equation}

 Choosing parameters such that this potential is bound from below, including $\lambda_{\eta_1}$, $\lambda_{\eta_2}$, $\lambda_{\chi_1}$, $\lambda_{\chi_2}$, $\lambda_{\eta_{1}\eta_{2}}$, $\lambda_{\eta_{1}\chi_{2}}$, $\lambda_{\chi_{1}\eta_{2}}$, $\lambda_{\chi_{1}\chi_{2}}>0$, there are four global minima given by $\eta_{1}=\pm{}v_1$, $\eta_{2}=\pm{}v_2$, $\chi_{1}=\chi_{2}=0$. Furthermore, we require that $\lambda_{\eta_{1}\chi_{1}}v^{2}_{1}>\mu^2_{\chi_{1}}$ and $\lambda_{\eta_{2}\chi_{2}}v^{2}_{2}>\mu^2_{\chi_{2}}$ to ensure that $\chi_{1}$ and $\chi_{2}$ attain tachyonic masses in the interiors of of the respective walls generated by $\eta_{1}$ and $\eta_{2}$. This ensure solutions where $\chi_{1}$ and $\chi_{2}$ form lump-like profiles are the most stable ones, analogously to the single kink-lump case. We also make the parameter choices $\mu^2_{\chi_{1}}>\lambda_{\chi_{1}\eta_{2}}v^2_2$ and $\mu^2_{\chi_{2}}>\lambda_{\eta_{1}\chi_{2}}v^2_1$. We choose the last two conditions so that $\chi_{1}$ does not condense along the edges at infinity along which the $\eta_{2}-\chi_{2}$ kink-lump solution interpolates and vice versa.

\begin{figure}[h]
\includegraphics[scale=0.65]{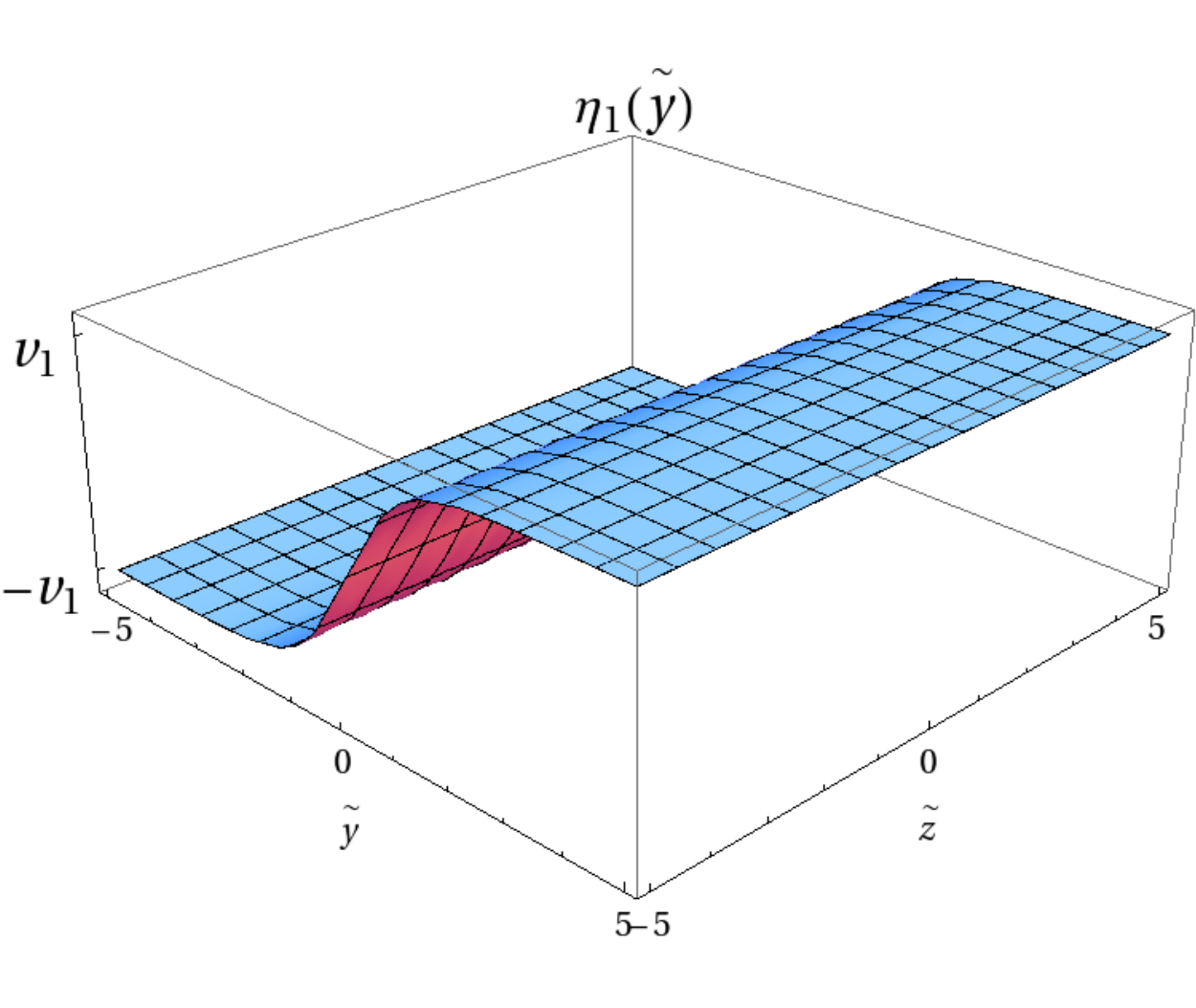}
\caption{A plot of $\eta_{1}$ for the solution in Eq.\ \ref{eq:perpendicularsolution}}
\label{fig:eta1plot}
\end{figure}

\begin{figure}[h]
\includegraphics[scale=0.65]{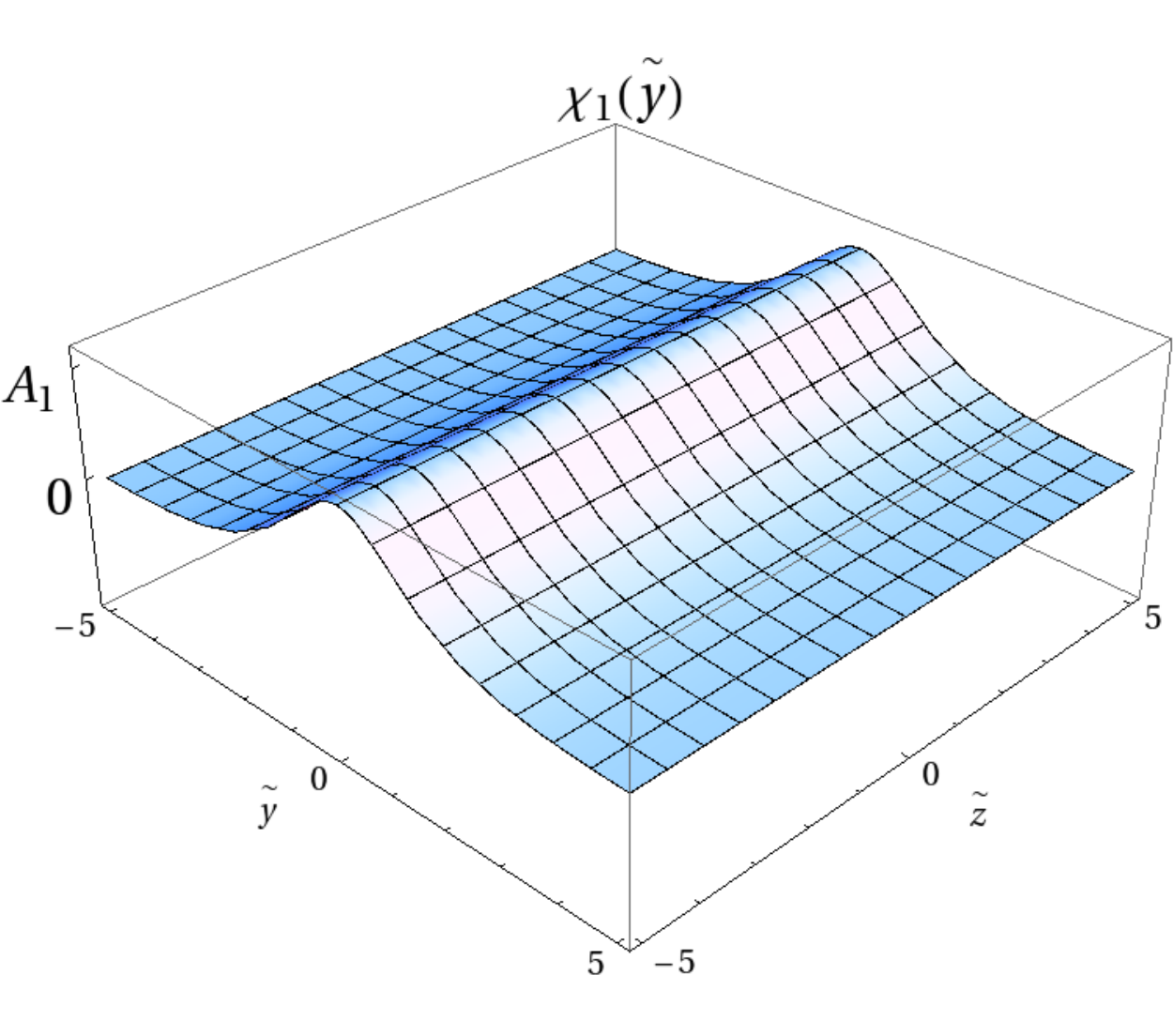}
\caption{A plot of $\chi_{1}$ for the solution in Eq.\ \ref{eq:perpendicularsolution}}
\label{fig:chi1plot}
\end{figure}

\begin{figure}[h]
\includegraphics[scale=0.65]{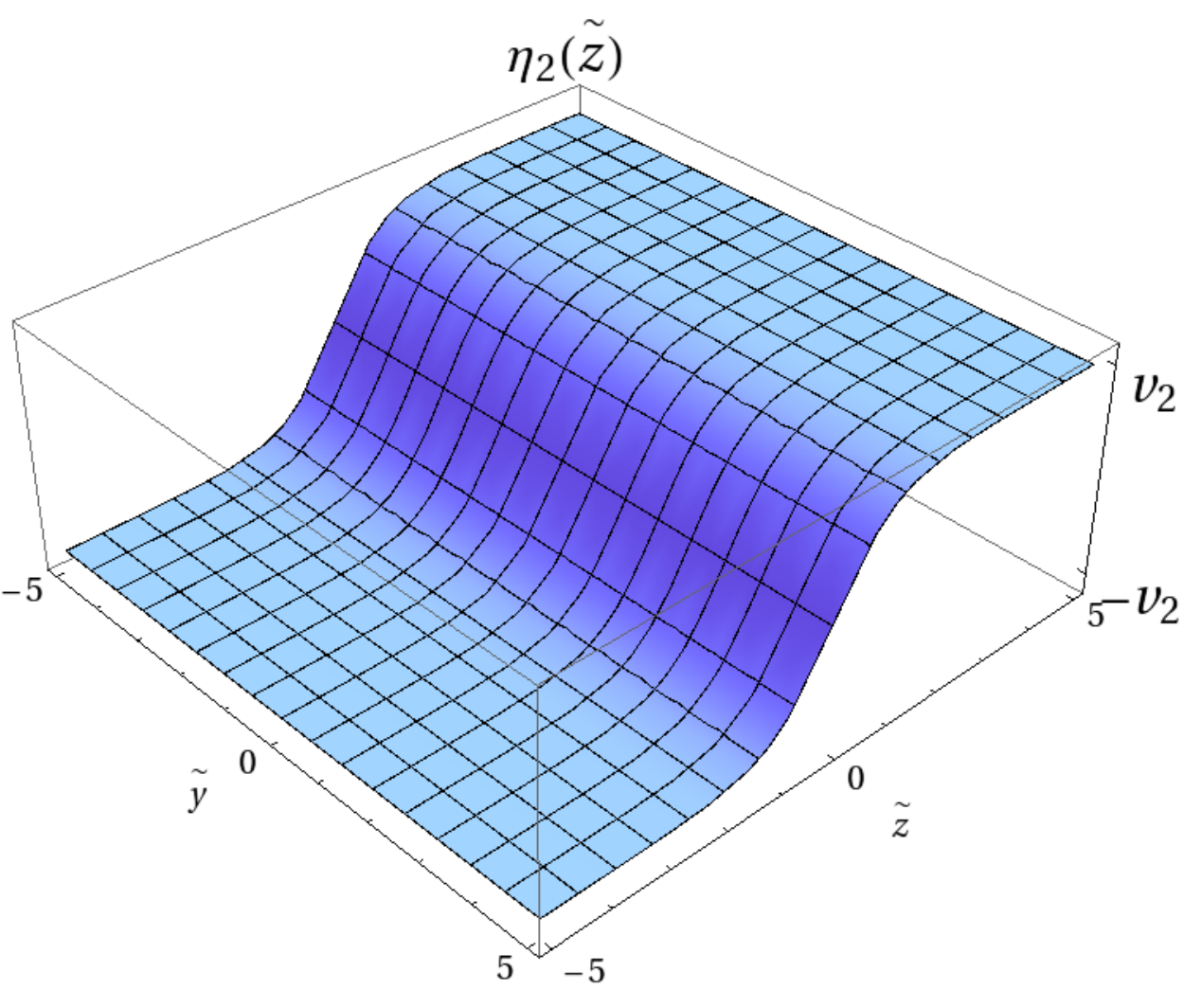}
\caption{A plot of $\eta_{2}$ for the solution in Eq.\ \ref{eq:perpendicularsolution}}
\label{fig:eta2plot}
\end{figure}

\begin{figure}[h]
\includegraphics[scale=0.65]{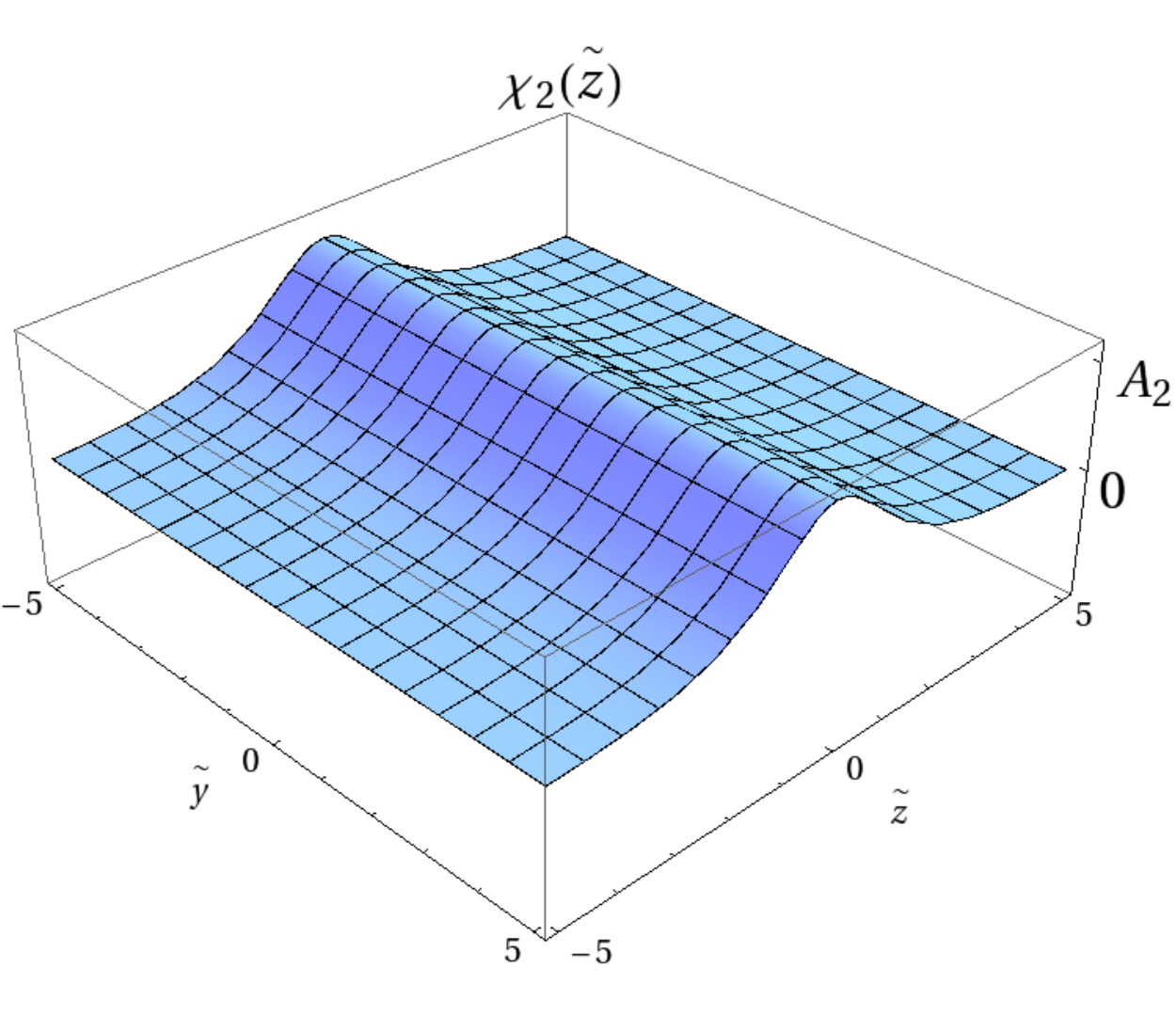}
\caption{A plot of $\chi_{2}$ for the solution in Eq.\ \ref{eq:perpendicularsolution}}
\label{fig:chi2plot}
\end{figure}

 To set up an intersecting domain wall solution, we must find a static solution for all of the four fields $\eta_1$, $\eta_2$, $\chi_1$ and $\chi_2$ to the Euler-Lagrange equations which at the very least interpolate amongst the four vacua at infinity along the corners of each quadrant in the $y-z$ plane. Firstly, we attempt to find a solution in which the walls are mutually perpendicular. Along one edge where $\eta_1$ is constant at one of its vacua $\pm{}v_1$ and where also $\chi_1=0$, the field $\eta_{2}$ should interpolate between the values $\pm{}v_2$ and $\chi_{2}$ should be zero at infinity along the edge and condense in the middle of the edge, much like the one-dimensional domain wall with internal structure discussed in the previous section. The same should apply to $\eta_{1}$ and $\chi_{1}$ along the edges where $\eta_{2}$ and $\chi_2$ are fixed. This motivates us to look for solutions obeying boundary conditions of the type
\begin{equation}
\begin{aligned}
\label{eq:perpendicularsolboundaryconditions}
&\eta_{1}(y=\pm{}\infty{}, z)=\pm{}v_1, \,{} \eta_{1}(y, z=\pm{}\infty{})=v_1\tanh{(ky)}, \\
&\eta_{2}(y=\pm{}\infty{}, z)=v_2\tanh{(lz)}, \,{} \eta_{2}(y, z=\pm{}\infty{})=\pm{}v_2, \\
&\chi_{1}(y=\pm{}\infty{}, z)=0, \,{} \chi_{1}(y, z=\pm{}\infty{})=A_{1}\sech{(ky)}, \\
&\chi_{2}(y=\pm{}\infty{}, z)=A_{2}\sech{(lz)}, \,{} \chi_{2}(y, z=\pm{}\infty{}) = 0.
\end{aligned}
\end{equation}
By making the parameter choice,
\begin{equation}
\begin{gathered}
\label{eq:parameterconditionsforasol}
\lambda_{\eta_1\eta_2\chi_2} = \lambda_{\chi_1\eta_2\chi_2} = \lambda_{\eta_1\chi_1\eta_2} = \lambda_{\eta_1\chi_1\chi_2} = \lambda_{\eta_1\chi_1\eta_2\chi_2} = 0, \\
g_{\eta_1\chi_1}=h_{\eta_1\chi_1}=g_{\eta_2\chi_2}=h_{\eta_2\chi_2}=0, \\
\lambda_{\eta_1\eta_2}v^2_1=\lambda_{\chi_1\eta_2}A^2_{1}, \quad{} \lambda_{\eta_1\eta_2}v^2_2=\lambda_{\eta_1\chi_2}A^2_{2}, \\
\lambda_{\eta_1\chi_2}v^2_1=\lambda_{\chi_1\chi_2}A^2_{1}, \quad{} \lambda_{\chi_1\eta_2}v^2_2=\lambda_{\chi_1\chi_2}A^2_{2}, \\
2\mu^2_{\chi_1}(\lambda_{\eta_1\chi_1}-\lambda_{\chi_1})+(\lambda_{\eta_1}\lambda_{\chi_1}-\lambda^2_{\eta_1\chi_1})v^2=0, \\
2\mu^2_{\chi_2}(\lambda_{\eta_2\chi_2}-\lambda_{\chi_2})+(\lambda_{\eta_2}\lambda_{\chi_2}-\lambda^2_{\eta_2\chi_2})v^2=0,
\end{gathered}
\end{equation}
one finds that
\begin{equation}
\begin{gathered}
\label{eq:perpendicularsolution}
\eta_{1}(y) = v_{1}\tanh{(ky)}, \quad{} \chi_{1}(y) = A_{1}\sech{(ky)}, \\
\eta_{2}(z) = v_{2}\tanh{(lz)}, \quad{} \chi_{2}(z) = A_{2}\sech{(lz)}, \\
\end{gathered}
\end{equation}
where $k^2=\mu^2_{\chi_1}$, $l^2=\mu^2_{\chi_2}$, $A^2_{1} = \frac{\lambda_{\eta_1\chi_1}v^2_1-2\mu^2_{\chi_1}}{\lambda_{\chi_1}}$, $A^2_{2} = \frac{\lambda_{\eta_2\chi_2}v^2_2-2\mu^2_{\chi_2}}{\lambda_{\chi_2}}$, is a solution to the four coupled Euler-Lagrange equations resulting from the potential in Eq.\ \ref{eq:scalarpotential} and satisfies the boundary conditions in Eq.\ \ref{eq:perpendicularsolboundaryconditions}. We give plots for $\eta_{1}$, $\chi_{1}$, $\eta_{2}$ and $\chi_{2}$ for the solution in Eq.\ \ref{eq:perpendicularsolution} in Fig.\ \ref{fig:eta1plot}, Fig.\ \ref{fig:chi1plot}, Fig.\ \ref{fig:eta2plot} and Fig.\ \ref{fig:chi2plot} respectively, in terms of the non-dimensionalized coordinates $\tilde{y}=ky$ and $\tilde{z}=lz$.

 It is important to note that there are other solutions with two kink-lump pairs to the coupled Euler-Lagrange equations. Without loss of generality, assume that $\eta_{1}$ and $\chi_{1}$ are the same as in Eq.\ \ref{eq:perpendicularsolution} but consider instead that $\eta_{2}$ and $\chi_{2}$ take the form
\begin{equation}
\begin{aligned}
\label{eq:angledsolution}
\eta_{2}(y,z) &= v_{2}\tanh{[l(\cos{\theta}y+\sin{\theta}z)]}, \\
\chi_{2}(y,z) &= A_{2}\sech{[l(\cos{\theta}y+\sin{\theta}z)]}.
\end{aligned}
\end{equation}
 Acting with the d'Alembertian operator on $\eta_{2}$ and $\chi_{2}$ yields 
\begin{equation}
\begin{aligned}
\label{eq:kineticangledkinklump}
\Box{}\eta_2 &= \frac{2l^2}{v^2_2}(\eta^2_2-v^2_2)\eta_{2}, \\
\Box{}\chi_2 &= \frac{2l^2}{A^2_{2}}\chi^3_{2}-l^2\chi_{2}.
\end{aligned}
\end{equation}

The resulting relations of the kinetic terms of $\eta_{2}$ and $\chi_{2}$ to themselves given in Eq.\ \ref{eq:kineticangledkinklump} are independent of the relative angle, $\theta$, between this kink-lump pair and that for $\eta_{1}$ and $\chi_{1}$. Hence, under the same parameter choice as in Eq.\ \ref{eq:parameterconditionsforasol}, 
\begin{equation}
\begin{aligned}
\label{eq:fullangledsolution}
\eta_{1}(y) &= v_{1}\tanh{(ky)},  \\
\chi_{1}(y) &= A_{1}\sech{(ky)}, \\
\eta_{2}(y,z) &= v_{2}\tanh{[l(\cos{\theta}y+\sin{\theta}z)]}, \\
\chi_{2}(y,z) &= A_{2}\sech{[l(\cos{\theta}y+\sin{\theta}z)]},
\end{aligned}
\end{equation}
is a solution to the Euler-Lagrange equations for general $\theta$. This means we have a whole class of solutions ranging from a solution in which the two walls are parallel ($\theta=0$), through intermediate angles of intersection, to the perpendicular solution of Eq.\ \ref{eq:perpendicularsolution}. The parallel and angled solutions obviously do not satisfy the same boundary conditions as the perpendicular solution given in Eq.\ \ref{eq:perpendicularsolboundaryconditions}. The angled solutions still divide the $y-z$ plane into four domains which tend to all four of the discrete global minima out at infinity. The parallel solution is a wall between two of these discrete vacua out at infinity along one direction. In calculating the energy density, the kinetic terms and the potentials describing the self-interactions and mass terms involving only $\eta_{1}$ and $\chi_{1}$ and likewise only $\eta_{2}$ and $\chi_{2}$ are associated with the energy density of the single kink-lump pairs, and integrating them over 
the directions normal to the walls yields the tensions associated with each kink-lump pair. In general, in theories with multiple domain-walls, the quartic interactions between the fields yielding different domain walls lead to a tension associated with the intersection or junction of those walls. In this scalar field theory and with the parameter choices we made in Eq.\ \ref{eq:parameterconditionsforasol}, we find that this tension is precisely zero,
\begin{equation}
\begin{aligned}
\label{eq:interbranetensions}
\varepsilon_{int} &= \frac{1}{2}\lambda_{\eta_1\eta_2}(\eta^2_1-v^2_1)(\eta^2_2-v^2_2)+\frac{1}{2}\lambda_{\eta_1\chi_2}(\eta^2_1-v^2_1)\chi^2_2 \\
                  &+\frac{1}{2}\lambda_{\chi_1\eta_2}\chi^2_1(\eta^2_2-v^2_2)+\frac{1}{2}\lambda_{\chi_1\chi_2}\chi^2_1\chi^2_2, \\
                  &= \frac{1}{2}\lambda_{\eta_1\eta_2}v^2_1v^2_2\sech^2{(ky)}\sech^2{(lu)} \\
                  &-\frac{1}{2}\lambda_{\eta_1\chi_2}v^2_1A^2_2\sech^2{(ky)}\sech^2{(lu)} \\
                  &-\frac{1}{2}\lambda_{\chi_1\eta_2}A^2_1v^2_2\sech^2{(ky)}\sech^2{(lu)} \\
                  &+\frac{1}{2}\lambda_{\chi_1\chi_2}A^2_1A^2_2\sech^2{(ky)}\sech^2{(lu)}, \\
                  &=0,
\end{aligned}
\end{equation}
for all angles $\theta$ where here we have used $u=\cos{\theta}y+\sin{\theta}z$. These solutions are degenerate in energy.

 It turns out that despite the energy degeneracy, neither the perpendicular nor angled solutions can evolve into the parallel solution. This is not surprising as the former two interpolate amongst the four vacua along the boundary, while the latter only interpolates between two of them so we expect them to be topologically distinct. To be precise, there exists a topological charge associated with the 2-dimensional boundary of the $y-z$ plane. The associated topological current is defined by
\begin{equation}
\label{eq:intersectingbranetopologicalcurrent}
J^{MNOP} = \epsilon^{MNOPQR}\epsilon^{ij}\partial_{Q}\eta_{i}\partial_{R}\eta_{j}.
\end{equation}
Clearly, the 6-divergence of this current vanishes and is thus conserved. The conserved topological charge associated with this current is just
\begin{equation}
\label{eq:intersectingbranetopologicalcurrent}
Q^{ABC} = \int{}d^6x J^{0ABC}.
\end{equation}
Since the background fields are solely dependent on $y$ and $z$, only the elements $Q^{ijk}$ where $i,j,k=1,2,3$ are non-zero. One can show that these charges are proportional to the integral
\begin{equation}
\label{eq:topchargeintegral}
I = \int_{\Sigma}dydz \big(\partial_{4}\eta_{1}\partial_{5}\eta_{2}-\partial_{5}\eta_1\partial_{4}\eta_{2}\big),
\end{equation}
where $\Sigma$ denotes the $y-z$ plane. Using Stokes' theorem, one can also write this as
\begin{equation}
\begin{aligned}
\label{eq:topchargelineintegral}
I &= \int_{\partial{}\Sigma} \eta_{1}\partial_{4}\eta_{2}dy+\eta_1\partial_{5}\eta_{2}dz, \\
  &= \int_{\partial{}\Sigma} \eta_{1}\nabla{}\eta_{2}.\mathbf{dl}.
\end{aligned}
\end{equation}

 One can easily show from Eqs.\ \ref{eq:topchargeintegral} and \ref{eq:topchargelineintegral} that for the perpendicular and angled solutions $I=4v_1v_2$ whereas for the parallel solution $I=0$. Thus the perpendicular and angled solutions are stable against decay or evolution to the parallel solution despite the energy degeneracy.

 The topological charge in Eq.\ \ref{eq:intersectingbranetopologicalcurrent} does not differentiate between the perpendicular and angled solutions. It is not clear whether this means that these solutions are in the same topological class as there could exist other topological charges which differ between the two types of solution. In case they are topologically equivalent, one can imagine that one could perform a small perturbation from the parameter region considered in generating the analytic solutions to ensure that the perpendicular solution is the most energetically favorable one since the energy degeneracy between the solutions is likely not true in general. From a rough glance at the interactions in Eq.\ \ref{eq:interbranetensions} one can see that if one performs a perturbation $\lambda_{\eta_{1}\eta_{2}}\rightarrow{}\lambda_{\eta_{1}\eta_{2}}+\epsilon$, with $\epsilon>0$, that the contribution $\epsilon{}\eta^2_{1}\eta^2_{2}$ is minimized for $\theta=90^{\circ}$ and tends towards infinity as $\theta{
}\rightarrow{}0$. Unfortunately, there are also resultant perturbations to the fields $\eta_{1}$, $\eta_{2}$, $\chi_{1}$ and $\chi_{2}$, and these perturbations satisfy four coupled partial differential equations which can only be solved numerically, so we can not give a definitive answer here.

 We also need to perform a local stability analysis of the solutions given in this section. Likewise, this requires numerically solving four non-linear coupled partial differential equations and we defer this to a later study. For the rest of this paper, we will assume that the perpendicular solution is stable and that one can always choose this to be the background solution. 

\section{Fermion Localization}
\label{sec:fermionlocalization}

 In this section, we show that fermions can be localized to the intersection of the two domain walls. Normally, in the case of a single domain wall, one localizes fermions to the centre by Yukawa coupling them to the relevant scalar field. In the case of a single domain wall with a Yukawa coupling of the form $\overline{\Psi}\Psi{}\eta$, as $\eta\rightarrow-\eta$ under the discrete $\mathbb{Z}_2$ symmetry, to preserve the symmetry and maintain topological stability, we require that the Dirac bilinear $\overline{\Psi}\Psi{}\rightarrow-\overline{\Psi}\Psi{}$ under the symmetry. This can be achieved by choosing the fermionic fields to transform individually as $\Psi\rightarrow{}\pm{}i\Gamma{}\Psi$, where $\Gamma$ is the gamma matrix associated with the direction parametrizing the profile of the domain-wall solution. In 4+1-dimensional (4+1D) theory, one often chooses $\Gamma=\Gamma^4=-i\gamma^5$ for example. 

 In six dimensions, with our background set-up, if one wishes to localize chiral fermions to the intersection region of the background solution, one must Yukawa couple the desired fermionic fields to all four scalar fields. However, with two independent $\mathbb{Z}_2$ symmetries, one has a problem in attempting to couple a fermionic field to both defects since if we perform the first $\mathbb{Z}_{2}$ transformation $\eta_1\rightarrow-\eta_1$, $\chi_1\rightarrow-\chi_1$, $\eta_2\rightarrow\eta_2$, $\chi_2\rightarrow\chi_2$, and maintain that $\overline{\Psi}\Psi\rightarrow-\overline{\Psi}\Psi$ under such a transformation, then $\overline{\Psi}\Psi\eta_2$ and $\overline{\Psi}\Psi\chi_2$ are not invariant under the first symmetry, and likewise $\overline{\Psi}\Psi\eta_1$ and $\overline{\Psi}\Psi\chi_1$ won't be under the second $\mathbb{Z}_2$ symmetry. Hence, we cannot localize fermions to the defect by using just scalar Yukawa interactions to all fields without compromising topological stability. This reflects the fact that we must choose our effective $6D$ localization bulk mass matrix carefully in order to localize chiral fermions on the intersection, an issue that was first raised in \cite{flachimasato6dintbrane}.

 However, in 5+1D there is another possibility, since in spacetimes of even dimensionality there always exists a chirality operator and thus there always exists a pseudoscalar bilinear in these spacetime dimensionalities. In 5+1D, the chirality operator $\Gamma^7$ is defined
\begin{equation}
\begin{aligned}
\label{eq:6dchiralityop}
\Gamma^7&=\Gamma^0\Gamma^1\Gamma^2\Gamma^3\Gamma^4\Gamma^5, \\
        &=\frac{1}{6!}\epsilon_{MNOPQR}\Gamma^M\Gamma^N\Gamma^O\Gamma^P\Gamma^Q\Gamma^R.
\end{aligned}
\end{equation}
One can then define the pseudoscalar bilinear $\overline{\Psi}\Gamma^7\Psi$. Now we consider the Yukawa terms
\begin{equation}
\begin{aligned}
\label{eq:fermionlocalizationpotential}
\mathcal{L}_{Yuk} &= -ih_{\eta_1}\overline{\Psi}\Gamma^7\Psi\eta_1-ih_{\chi_1}\overline{\Psi}\Gamma^7\Psi\chi_1 \\
                  &+h_{\eta_2}\overline{\Psi}\Psi\eta_2+h_{\chi_2}\overline{\Psi}\Psi\chi_2,
\end{aligned}
\end{equation}
and ask if it is possible to define two independent transformations for each $\mathbb{Z}_2$ symmetry for $\Psi$ such that for the first symmetry in which $\eta_1\rightarrow-\eta_1$ and $\chi_1\rightarrow-\chi_1$ we have $\overline{\Psi}\Gamma^7\Psi\rightarrow{}-\overline{\Psi}\Gamma^7\Psi$ but $\overline{\Psi}\Psi\rightarrow\overline{\Psi}\Psi$, while for the second reflection symmetry $\eta_2\rightarrow-\eta_2$ and $\chi_2\rightarrow-\chi_2$ we have $\overline{\Psi}\Gamma^7\Psi$ unchanged but $\overline{\Psi}\Psi\rightarrow-\overline{\Psi}\Psi$. Due to the fact that $\Gamma^7$ anticommutes with the gamma matrices, for the second $\mathbb{Z}_{2}$ one can easily show that the usual type of transformation $\Psi\rightarrow{}i\Gamma^5\Psi$ can be chosen. For the first $\mathbb{Z}_{2}$, one can show that the transformation $\Psi\rightarrow{}i\Gamma^4\Gamma^7\Psi$ induces the transformation $\overline{\Psi}\Psi\rightarrow\overline{\Psi}\Psi$ and $\overline{\Psi}\Gamma^7\Psi\rightarrow-\overline{\Psi}\Gamma^7\Psi$. 
Hence, we have shown that there exists a mechanism to couple a fermionic field to all four background scalar fields with the combination of scalar and pseudoscalar Yukawa couplings given in Eq.\ \ref{eq:fermionlocalizationpotential}.

 Henceforth, we assume boundary conditions such that we have the perpendicular intersecting domain-wall solution and we take the following sets of transformations to be our reflection symmetries which ensure topological stability of the background,
\begin{equation}
\begin{aligned}
\label{eq:reflectionsym1}
y&\rightarrow{}-y, \\
z&\rightarrow{}z,   \\
\eta_1&\rightarrow{}-\eta_1, \\
\chi_1&\rightarrow{}-\chi_1, \\
\eta_2&\rightarrow{}\eta_2,  \\
\chi_2&\rightarrow{}\chi_2, \\
\Psi{}&\rightarrow{}i\Gamma^4\Gamma^7\Psi, 
\end{aligned}
\end{equation}
and
\begin{equation}
\begin{aligned}
\label{eq:reflectionsym1}
y&\rightarrow{}y, \\
z&\rightarrow{}-z,   \\
\eta_1&\rightarrow{}\eta_1, \\
\chi_1&\rightarrow{}\chi_1, \\
\eta_2&\rightarrow{}-\eta_2,  \\
\chi_2&\rightarrow{}-\chi_2, \\
\Psi{}&\rightarrow{}i\Gamma^5\Psi. 
\end{aligned}
\end{equation}
 
 We now need to show that there is indeed a chiral zero mode localized to the intersection of the domain walls. Writing down the resultant 6D Dirac equation for $\Psi$, we have
\begin{equation}
\label{eq:6ddiracequation}
i\Gamma^M\partial_{M}\Psi+iW_{1}(y)\Gamma^7\Psi-W_{2}(z)\Psi=0,
\end{equation}
where
\begin{equation}
\begin{aligned}
\label{eq:6dyukawasuperpotentials}
W_{1}(y) &= h_{\eta_1}\eta_{1}(y)+h_{\chi_1}\chi_{1}(y), \\
W_{2}(z) &= h_{\eta_2}\eta_{2}(z)+h_{\chi_2}\chi_{2}(z).
\end{aligned}
\end{equation}

 In order to perform dimensional reduction and calculate the profiles of the modes of $\Psi$, we must choose a basis for the 5+1D Clifford algebra. One can show that 
\begin{equation}
\begin{aligned}
\label{eq:6dgammamatrices}
\Gamma^{\mu} &= \sigma_{1}\otimes{}\gamma^{\mu} = \begin{pmatrix} 0 && \gamma^{\mu} \\
                                                                 \gamma^{\mu} && 0 \end{pmatrix}, \\
\Gamma^4 &= \sigma_{1}\otimes{}-i\gamma^{5} = \begin{pmatrix} 0 && -i\gamma^{5} \\
                                                                 -i\gamma^{5} && 0 \end{pmatrix}, \\
\Gamma^5 &= -i\sigma_{3}\otimes{}\mathbb{1} = \begin{pmatrix} -i && 0 \\
                                                                        0 && i \end{pmatrix},
\end{aligned}
\end{equation}
satisfies the 5+1D Clifford algebra
\begin{equation}
\label{eq:6dcliffordalgebra}
\{\Gamma^M, \Gamma^N\} = 2\eta^{MN},
\end{equation}
and is thus an appropriate choice of basis for the 5+1D gamma matrices. In this basis the 6D chirality operator is 
\begin{equation}
\label{eq:6dchiralityopinbasis}
\Gamma^7 = \sigma_{2}\otimes{}\mathbb{1} = \begin{pmatrix} 0 && i \\
                                                          -i && 0 \end{pmatrix}.
\end{equation}
Decomposing $\Psi$ into components $\Psi_{\pm}$ which have 4 complex components and are eigenvectors of $\Gamma^5$,
\begin{equation}
\label{eq:psifourcompspinors}
\Psi = \begin{pmatrix} \Psi_{+} \\
                       \Psi_{-} \end{pmatrix},
\end{equation}
one can shown that the 6D Dirac equation reduces to 
\begin{subequations}
\label{eq:6ddiracequationpsipmcomponents}
\begin{align}
(i\gamma^{\mu}\partial_{\mu}+\gamma^5\partial_{4}+W_{1}(y))\Psi_{+}-\partial_{5}\Psi_{-}-W_{2}(z)\Psi_{-}&=0, \label{eq:6ddiracequationpsipmcomponentsa} \\
(i\gamma^{\mu}\partial_{\mu}+\gamma^5\partial_{4}-W_{1}(y))\Psi_{-}+\partial_{5}\Psi_{+}-W_{2}(z)\Psi_{+}&=0. \label{eq:6ddiracequationpsipmcomponentsb}
\end{align}
\end{subequations}

 To calculate the profiles of all modes, due to the fact that the excited Kaluza-Klein (KK) modes are usually Dirac fermions it is useful to find the corresponding Klein-Gordon (KG) equation that the components $\Psi_{\pm}$ satisfy. Operating with $(i\gamma^{\mu}\partial_{\mu}+\gamma^5\partial_{4}-W_{1}(y))$ from the left on Eq.\ \ref{eq:6ddiracequationpsipmcomponentsa} and likewise $(i\gamma^{\mu}\partial_{\mu}+\gamma^5\partial_{4}+W_{1}(y))$ on Eq.\ \ref{eq:6ddiracequationpsipmcomponentsb}, one obtains the KG equations 
\begin{equation}
\label{eq:fermionkgequations}
\Box{}\Psi_{\pm}+(W_{1}(y)^2\mp{}W'_{1}(y)\gamma_5)\Psi_{\pm}+(W_{2}(z)^2\pm{}W'_{2}(z))\Psi_{\pm}=0.
\end{equation}
Now we expand each of $\Psi_{\pm}$ as a series of modes. As one can see from Eq.\ \ref{eq:6ddiracequationpsipmcomponents}, the $y$-dependent piece is chirality-dependent while the $z$-dependent piece is not. The z-dependent piece is only dependent on whether the component is $\Psi_{-}$ or $\Psi_{+}$. Thus we make the expansion
\begin{equation}
\label{eq:fermionkkexpansion} 
\begin{aligned}
\Psi_{\pm}&(x_{\mu}, y, z) = \\
&\sum_{\substack{m}}f^{m}_{\pm{}L}(y)g^{m}_{\pm{}}(z)\varphi^{m}_{\pm{}L}(x_{\mu})+f^{m}_{\pm{}R}(y)g^{m}_{\pm{}}(z)\varphi^{m}_{\pm{}R}(x_{\mu}). 
\end{aligned}
\end{equation}
Here, $m$ just denotes some mass eigenvalue, $\varphi^{m}_{\pm{}L,R}$ denotes the 3+1D left/right-chiral mode of mass $m$ embedded in the component $\Psi_{\pm}$ of $\Psi$, and $f^{m}_{\pm{}L,R}(y)$ and $g^{m}_{\pm{}}(z)$ are profiles for these modes along the $y$ and $z$ directions respectively. Since any 3+1D fermionic mode should satisfy a corresponding Klein-Gordon equation, let us substitute the expansion into Eq.\ \ref{eq:fermionkgequations} and demand that modes satisfy $\Box{}_{3+1}\varphi^{m}_{\pm{}L,R}(x_{\mu})=-m^2\varphi^{m}_{\pm{}L,R}(x_{\mu})$. We find that Eq.\ \ref{eq:fermionkgequations} reduces to
\begin{equation}
\label{eq:kgprofileeq}
\begin{aligned}
&\big[-\frac{d^2f^{m}_{\pm{}L}}{dy^2}+(W_{1}(y)^2\pm{}W'_{1}(y))f^{m}_{\pm{}L}(y))\big]g^{m}_{\pm{}}(z) \\
&+f^{m}_{\pm{}L}(y)\big[-\frac{d^2g^{m}_{\pm{}}}{dz^2}+(W_{2}(z)^2\pm{}W'_{2}(z))g^{m}_{\pm{}}(z)\big] \\
&=m^2f^{m}_{\pm{}L}(y)g^{m}_{\pm{}}(z), \\
&\big[-\frac{d^2f^{m}_{\pm{}R}}{dy^2}+(W_{1}(y)^2\mp{}W'_{1}(y))f^{m}_{\pm{}R}(y)\big]g^{m}_{\pm{}}(z) \\
&+f^{m}_{\pm{}R}(y)\big[-\frac{d^2g^{m}_{\pm{}}}{dz^2}+(W_{2}(z)^2\pm{}W'_{2}(z))g^{m}_{\pm{}}(z)\big] \\
&=m^2f^{m}_{\pm{}R}(y)g^{m}_{\pm{}}(z),
\end{aligned}
\end{equation}
for the left and right-chiral components respectively.

 Demanding that the profiles satisfy the following Schr$\ddot{o}$dinger equations (SE)
\begin{subequations}
\label{eq:fermionschrodingeq}
\begin{align}
-\frac{d^2f^{m}_{\pm{}L}}{dy^2}+(W_{1}(y)^2\pm{}W'_{1}(y))f^{m}_{\pm{}L}(y)=\lambda^1_{\pm{}L}f^{m}_{\pm{}L}(y),  \label{eq:fermionschrodingeqpmlefty}  \\
-\frac{d^2f^{m}_{\pm{}R}}{dy^2}+(W_{1}(y)^2\mp{}W'_{1}(y))f^{m}_{\pm{}R}(y)=\lambda^1_{\pm{}R}f^{m}_{\pm{}R}(y), \label{eq:fermionschrodingeqpmrighty}    \\
-\frac{d^2g^{m}_{\pm{}}}{dz^2}+(W_{2}(z)^2\pm{}W'_{2}(z))g^{m}_{\pm{}}(z)=\lambda^2_{\pm{}}g^{m}_{\pm{}}(z),  \label{eq:fermionschrodingeqpmz}
\end{align}
\end{subequations}
we find that the values of the squared masses of the localized KK modes are
\begin{equation}
\label{eq:fermionmasssqrgeneralminus}
m^2_{\pm{}L,R} = \lambda^1_{\pm{}L,R}+\lambda^2_{\pm{}},
\end{equation}
for the modes embedded in $\Psi_{\pm{}}$.

 Given the definitions for $W_{1}$ and $W_{2}$ in Eq.\ \ref{eq:6dyukawasuperpotentials}, one can see that the potentials of Eq.\ \ref{eq:fermionschrodingeq} are hyperbolic Scarf potentials. These potentials are well known and can be solved analytically \cite{castillohyperscarfpot, levaishapeinvariantpots, shapeinvkharesukdab}. For simplicity, let us assume that both $h_{\eta_1}$ and $h_{\eta_2}$ are positive definite. Non-dimensionalizing all variables and parameters except $m$ as 
\begin{equation}
\label{eq:fermionnondimparameters}
\begin{gathered}
\tilde{y}=ky, \qquad{} \tilde{z}=lz, \\
\tilde{h}_{\eta_{1}} = \frac{h_{\eta_1}v_1}{k}, \qquad{} \tilde{h}_{\chi_{1}} = \frac{h_{\chi_1}A_1}{k}, \\
\tilde{h}_{\eta_{2}} = \frac{h_{\eta_2}v_2}{l}, \qquad{} \tilde{h}_{\chi_{2}} = \frac{h_{\chi_2}A_2}{l}, \\
\tilde{\lambda}^1_{\pm{}L,R} = \frac{\lambda^1_{\pm{}L,R}}{k^2}, \qquad{} \tilde{\lambda}^2_{\pm{}} = \frac{\lambda^2_{\pm{}}}{l^2},
\end{gathered}
\end{equation}
and non-dimensionalizing the profiles as 
\begin{equation}
\label{eq:fermionprofilesnondim}
\begin{gathered}
\tilde{f}^{m}_{\pm{}L,R}(\tilde{y}) = k^{-\frac{1}{2}}f^{m}_{\pm{}L,R}(y), \\
\tilde{g}^{m}_{\pm{}}(\tilde{z}) = l^{-\frac{1}{2}}g^{m}_{\pm{}}(z),
\end{gathered}
\end{equation}
one can show that each of the Eqs.\ \ref{eq:fermionschrodingeqpmlefty}, \ref{eq:fermionschrodingeqpmrighty}, and \ref{eq:fermionschrodingeqpmz} has a finite number of localized, square-normalizable solutions as well as a delocalized continuum. First, let us start with the $z$-dependent equations. For positive $h_{\eta_2}$, it is the potential for $\Psi_{-}$, $\tilde{W}_{2}^2-\tilde{W}'_{2}$ which generates a series of $\lceil{}h_{\eta_2}\rceil{}$ localized modes starting from an eigenvalue of $\tilde{\lambda}^{2}_{-0}=0$. The eigenvalues of these modes are given as 
\begin{equation}
\label{eq:psiminuszhsevalues}
\tilde{\lambda}^2_{-n} = 2n\tilde{h}_{\eta_2}-n^2, \quad{} n=0, 1, ..., \lfloor{}\tilde{h}_{\eta_2}\rfloor{}.
\end{equation}
The $\tilde{\lambda}^2_{-0}$ profile is given by
\begin{equation}
\label{eq:zeromodezprofile}
\tilde{g}^{0}_{-}(\tilde{z}) = \tilde{D}^{0}_{-}e^{-\tilde{h}_{\eta_2}\log{[\cosh{(\tilde{z})}]}-2\tilde{h}_{\chi_2}\arctan{[\tanh{(\tilde{z}/2)}]}},
\end{equation}
and the profiles for the excited localized modes can be generated by applying the ladder operator which is proportional to $W_{2}(\tilde{z})-\frac{d}{d\tilde{z}}$. 
As for the potential for $\Psi_{+}$, $\tilde{W}_{2}^2+\tilde{W}'_{2}$, given our parameter choice there is no solution with $\tilde{\lambda}^2_{+0}=0$. Rather, there are $\lfloor{}\tilde{h}_{\eta_2}\rfloor{}$ localized modes starting from an eigenvalue $\tilde{\lambda}^2_{+1}=2\tilde{h}_{\eta_2}-1$ (provided $\tilde{h}_{\eta_2}>1)$. All this implies that, in considering just the interaction with the $z$-dependent kink-lump, for $\tilde{h}_{\eta_2}>0$, there is a massless 4+1D Dirac zero mode generated in $\Psi_{-}$ and then a tower of massive 4+1D Dirac modes embedded in both $\Psi_{-}$ and $\Psi_{+}$. In then further considering the interaction with the $y$-dependent part of the solution and calculating the $f_{\pm{}L,R}$, each of these 4+1D modes will generate a tower of 3+1D left- and right-chiral modes localized to the domain-wall intersection. Obviously, any chiral zero mode produced on the intersection must be embedded in the massless 4+1D zero mode of $\Psi_{-}$ in this case. If we choose $\tilde{h}_{\eta_2}$ to be negative instead, the roles of $\Psi_{-}$ and $\Psi_{+}$ in this situation are reversed, and the chiral zero mode must therefore be embedded in $\Psi_{+}$.

 Now let us analyse the $y$-dependent equations. These equations are also SE's with the hyperbolic Scarf potentials. In this case the particular form of the hyperbolic Scarf potential for the modes is dependent on chirality. Since we are assuming $\tilde{h}_{\eta_2}>0$ without loss of generality, let us focus on the left- and right-chiral modes embedded in $\Psi_{-}$ first, since as noted above this is the component containing any potential chiral zero mode. Looking at Eqs.\ref{eq:fermionschrodingeqpmlefty} and \ref{eq:fermionschrodingeqpmrighty}, one sees that the potentials for the left- and right-chiral components in $\Psi_{-}$ are $\tilde{W}_{1}(\tilde{y})^2-\tilde{W}'_{1}(\tilde{y})$ and $\tilde{W}_{1}(\tilde{y})^2+\tilde{W}'_{1}(\tilde{y})$ respectively. Hence, we easily deduce that for $\tilde{h}_{\eta_1}>0$, the equation for the left-chiral modes of $\Psi_{-}$ has the same form as that for the $z$-dependent profile equation for $\Psi_{-}$ and thus has a mode starting from an eigenvalue of $\tilde{\lambda}^1_{-L0}=0$, and likewise that for the right-chiral modes has the same form as that for the $\tilde{g}^m_{+}(\tilde{z})$ and thus only has solutions with positive definite eigenvalues. Since from Eq.\ \ref{eq:fermionmasssqrgeneralminus} we know that these eigenvalues directly contribute to the mass, this implies that for the choice $\tilde{h}_{\eta_{1}}>0$, $\tilde{h}_{\eta_{2}}>0$, there is a single massless left-chiral zero mode embedded in $\Psi_{-}$ localized to the intersection.

\begin{figure}[h]
\includegraphics[scale=0.65]{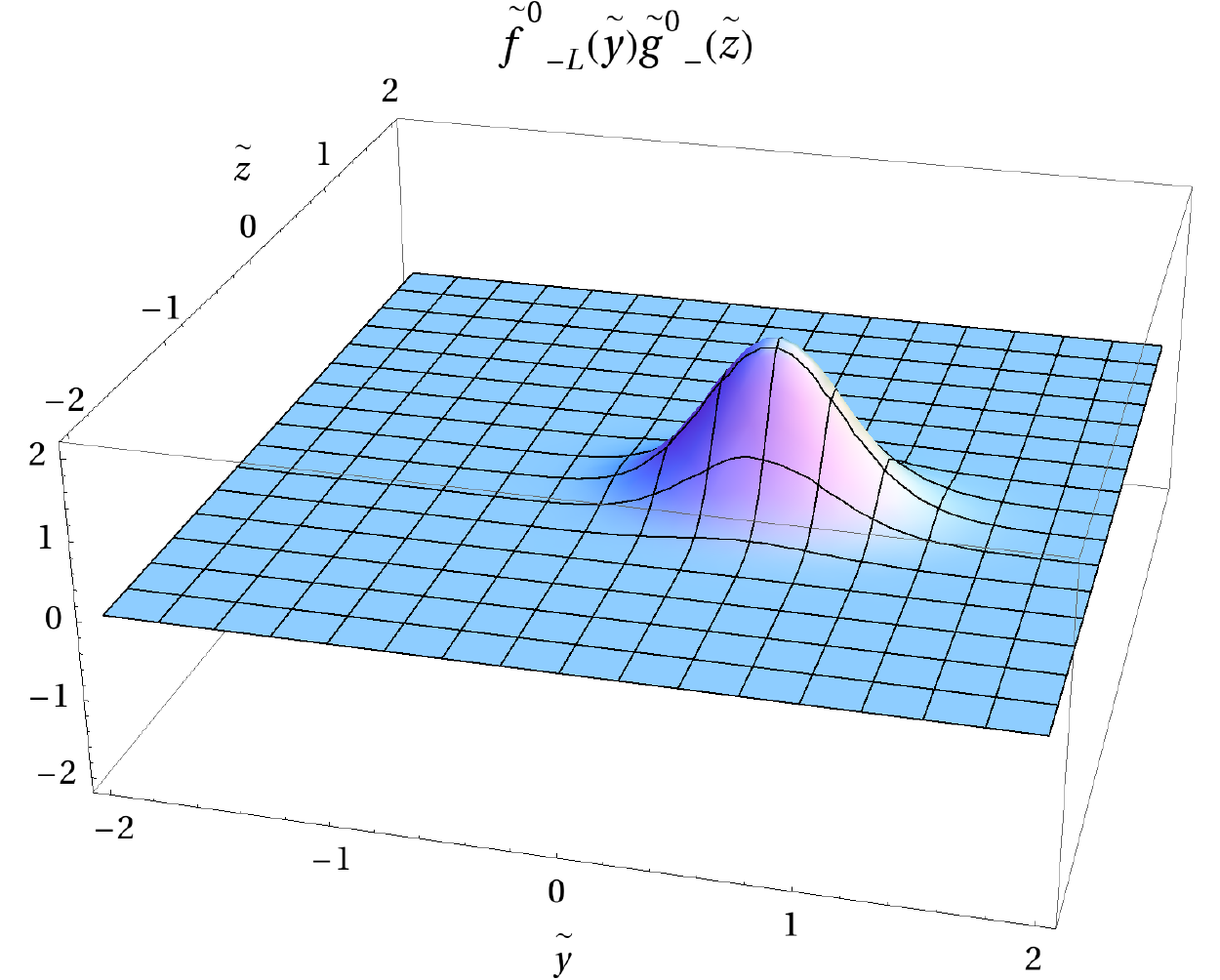}
\caption{A plot of the profile for the left chiral zero mode embedded in $\Psi_{-}$ for the parameter choice $\tilde{h}_{\eta_1}=10$, $\tilde{h}_{\chi_1}=-5$, $\tilde{h}_{\eta_2}=20$, and $\tilde{h}_{\chi_2}=4$}
\label{fig:chiralzeromode}
\end{figure}

\begin{figure}[h]
\includegraphics[scale=0.65]{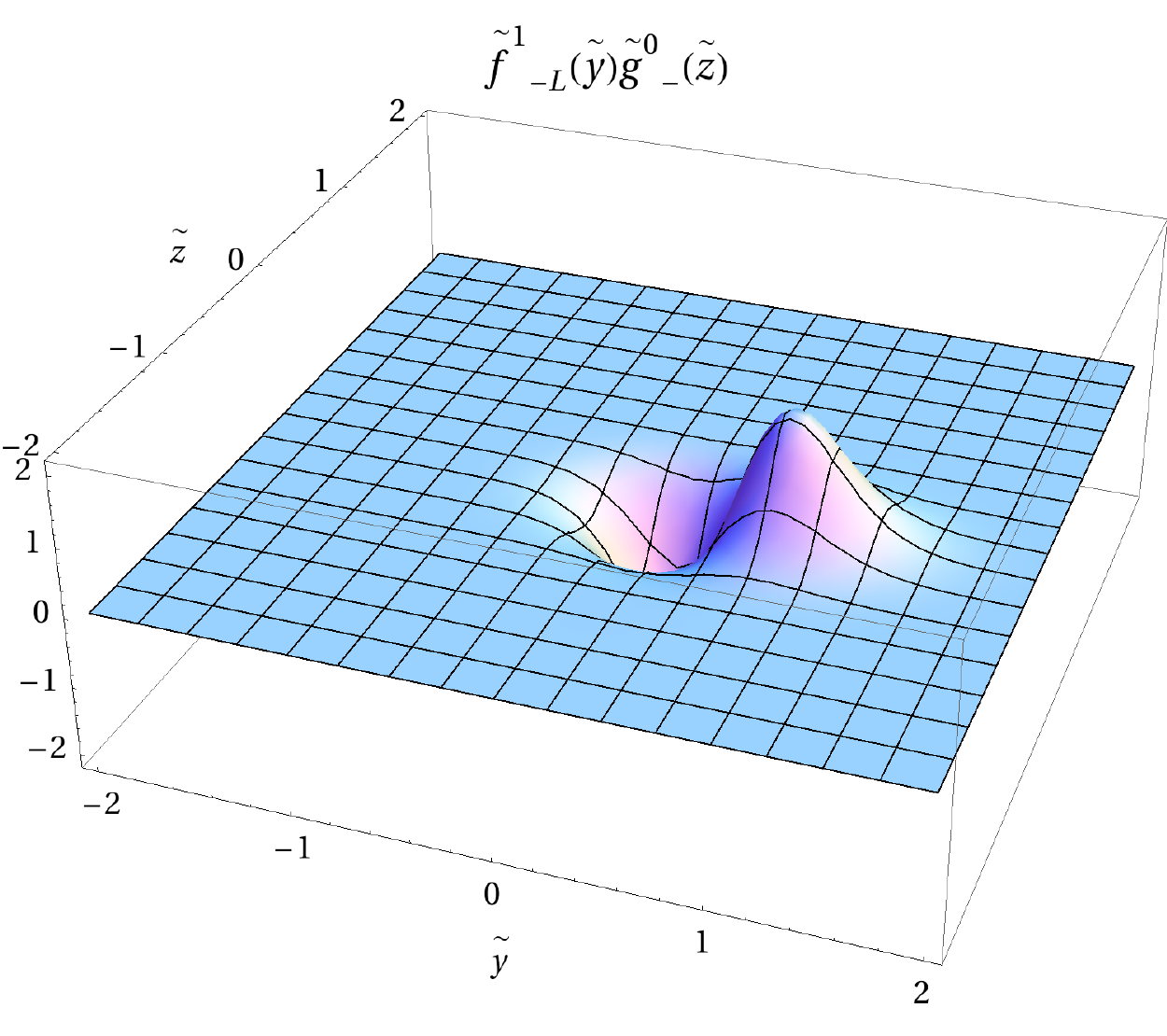}
\caption{A plot of the profile for the left-chiral $i=1$, $j=0$ mode in $\Psi_{-}$ for the parameter choice $\tilde{h}_{\eta_1}=10$, $\tilde{h}_{\chi_1}=-5$, $\tilde{h}_{\eta_2}=20$, and $\tilde{h}_{\chi_2}=4$}
\label{fig:fermionmode10}
\end{figure}

\begin{figure}[h]
\includegraphics[scale=0.65]{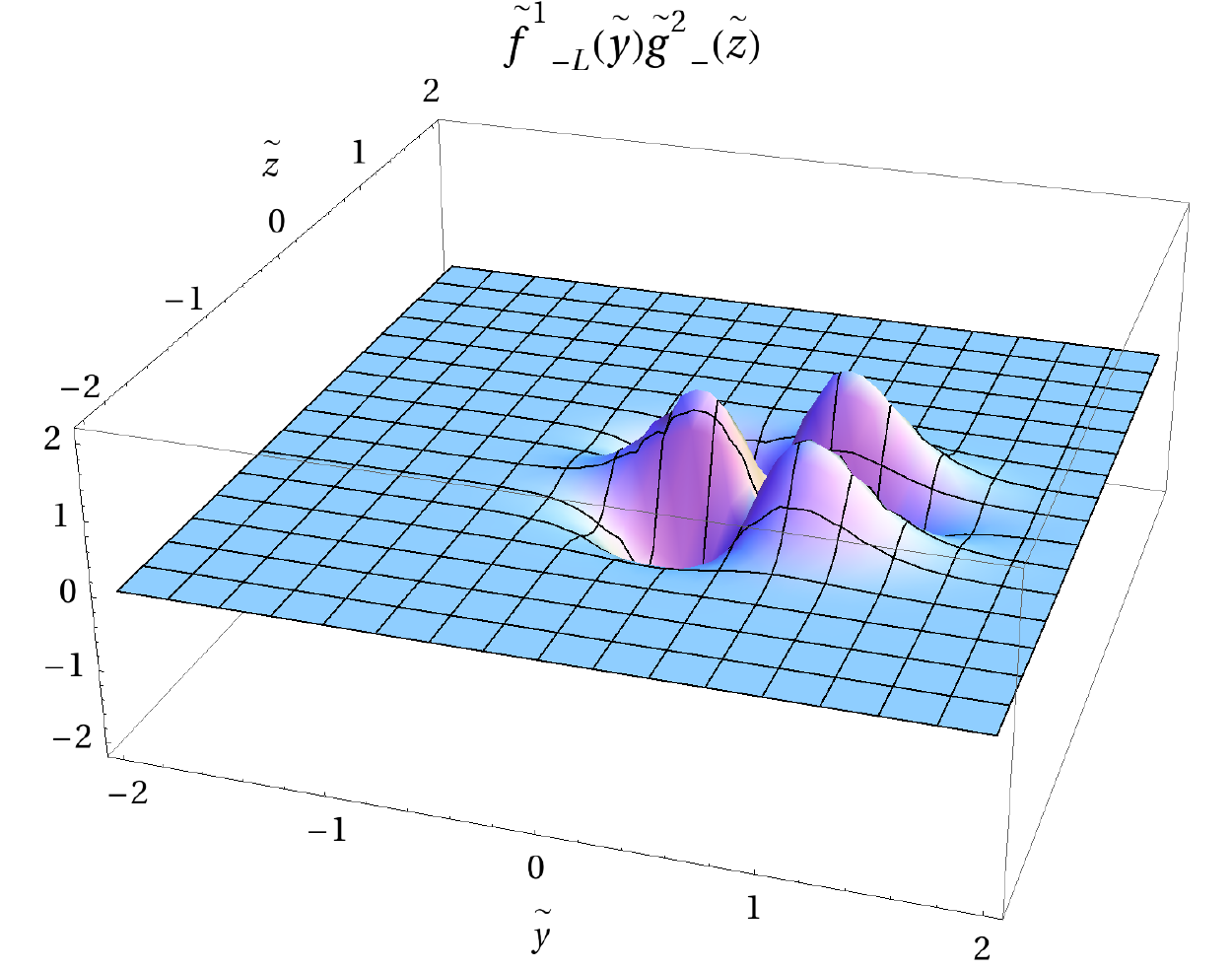}
\caption{A plot of the profile for the left-chiral $i=1$, $j=2$ mode in $\Psi_{-}$ for the parameter choice $\tilde{h}_{\eta_1}=10$, $\tilde{h}_{\chi_1}=-5$, $\tilde{h}_{\eta_2}=20$, and $\tilde{h}_{\chi_2}=4$}
\label{fig:fermionmode12}
\end{figure}

\begin{figure}[h]
\includegraphics[scale=0.65]{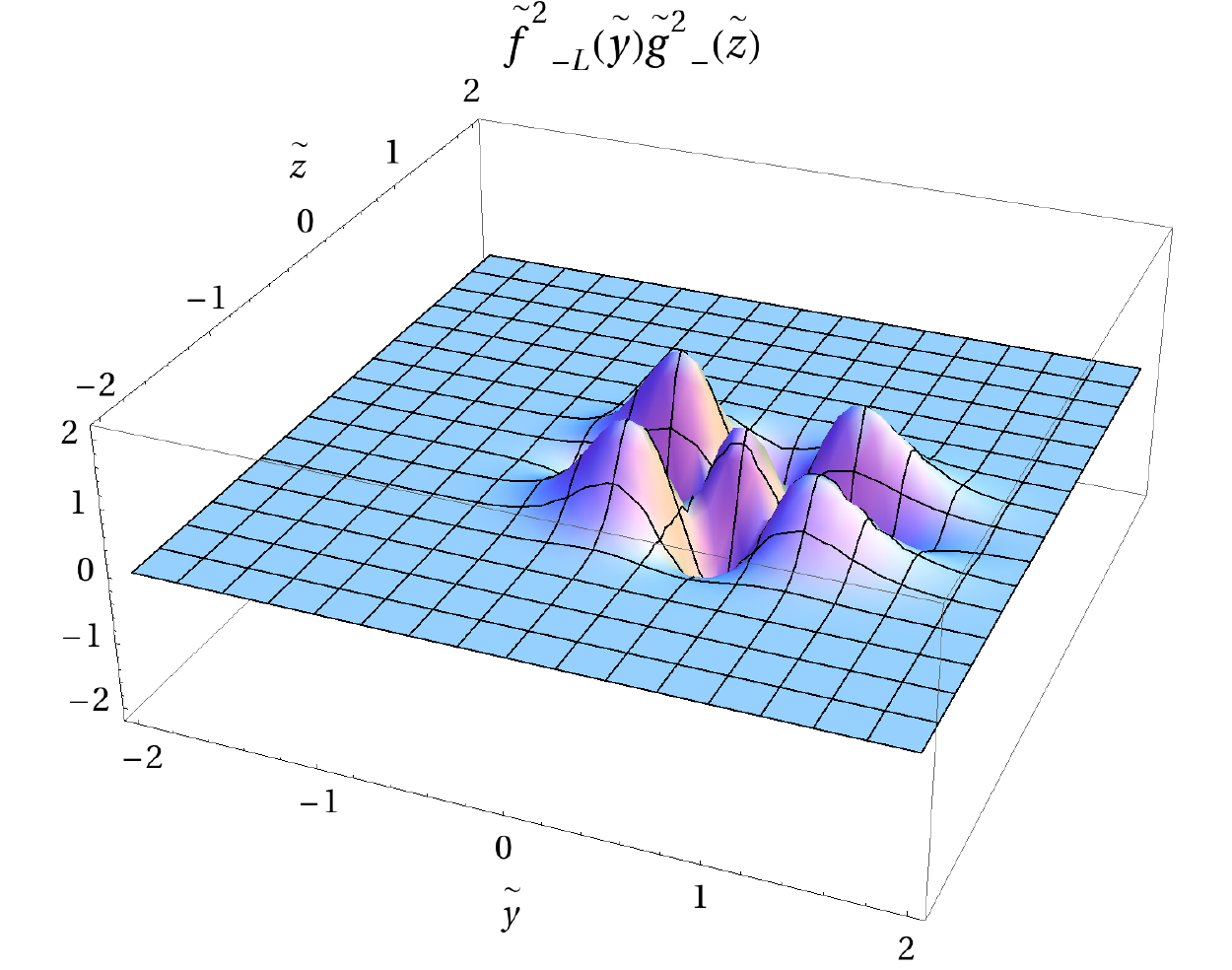}
\caption{A plot of the profile for the left-chiral $i=2$, $j=2$ mode in $\Psi_{-}$ for the parameter choice $\tilde{h}_{\eta_1}=10$, $\tilde{h}_{\chi_1}=-5$, $\tilde{h}_{\eta_2}=20$, and $\tilde{h}_{\chi_2}=4$}
\label{fig:fermionmode22}
\end{figure}

 The resultant solution for the $\lambda^1_{-L0}=0$ eigenfunction, $\tilde{f}^{0}_{-L}(\tilde{y})$ is given by 
\begin{equation}
\label{eq:zeromodeyprofile}
\tilde{f}^{0}_{-L}(\tilde{y}) = \tilde{C}^{0}_{-L}e^{-\tilde{h}_{\eta_1}\log{[\cosh{(\tilde{y})}]}-2\tilde{h}_{\chi_1}\arctan{[\tanh{(\tilde{y}/2)}]}},
\end{equation}
and thus the full profile over the $y-z$ plane for the left chiral zero mode is 
\begin{equation}
\label{eq:fullchiralzeromodeprofile}
\tilde{F}^{0}_{-L}(\tilde{y},\tilde{z}) = \tilde{f}^{0}_{-L}(\tilde{y})\tilde{g}^{0}_{-}(\tilde{z}).
\end{equation}
and a plot of this is shown in Fig.\ \ref{fig:chiralzeromode}.

 The $y$-dependent profiles for the higher localized left-chiral modes can be accessed by applying a ladder operator proportional to $W_{1}(\tilde{y})-\frac{d}{d\tilde{y}}$. The eigenvalues of these $\tilde{f}^{n}_{-L}$ profiles are given as 
\begin{equation}
\label{eq:minusleftfevalues}
\tilde{\lambda}^1_{-Ln} = 2\tilde{h}_{\eta_1}n-n^2.
\end{equation}
Putting Eqs.\ \ref{eq:psiminuszhsevalues} and \ref{eq:minusleftfevalues} together and converting back to dimensionful variables, we can see that the resultant squared masses of the left-chiral modes embedded in $\Psi_{-}$ are 
\begin{equation}
\begin{aligned}
\label{eq:minusleftkkmasses}
m^2_{-Lij} &= \tilde{\lambda}^1_{-Li}k^2+\tilde{\lambda}^2_{-j}l^2, \\
           &= 2ih_{\eta_1}v_{1}k-2i^2k^2+2jh_{\eta_2}v_{2}l-2j^2l^2, \\ 
           &i=0,1,..., \lfloor{}\tilde{h}_{\eta_1}\rfloor{}, \quad{}j=0,1,..., \lfloor{}\tilde{h}_{\eta_2}\rfloor{}.
\end{aligned}
\end{equation}
For the parameter choice $\tilde{h}_{\eta_1}=10$, $\tilde{h}_{\chi_1}=-5$, $\tilde{h}_{\eta_2}=20$, and $\tilde{h}_{\chi_2}=4$, we give a plot of the full profile $\tilde{F}^{0}_{-L}(\tilde{y},\tilde{z})$ for the zero mode ($i=0$, $j=0$) in Fig.\ \ref{fig:chiralzeromode}. We also give plots for the analogous profiles of the left-chiral $i=1$, $j=0$, $i=1$, $j=2$, and $i=2$, $j=2$ KK modes in Fig.\ \ref{fig:fermionmode10}, Fig.\ \ref{fig:fermionmode12} and Fig.\ \ref{fig:fermionmode22} respectively.

For the right-chiral modes of $\Psi_{-}$, there is no zero mode, but the massive modes have the same mass as the left-chiral counterparts, and the masses are 
\begin{equation}
\begin{aligned}
\label{eq:minusleftkkmasses}
m^2_{-Rij} &= \tilde{\lambda}^1_{-Ri}k^2+\tilde{\lambda}^2_{-j}l^2, \\
           &= 2ih_{\eta_1}v_{1}k-2i^2k^2+2jh_{\eta_2}v_{2}l-2j^2l^2, \\ 
           &i=1,..., \lfloor{}\tilde{h}_{\eta_1}\rfloor{}, \quad{}j=0,1,..., \lfloor{}\tilde{h}_{\eta_2}\rfloor{}.
\end{aligned}
\end{equation}
For the left-chiral and right-chiral localized modes embedded in $\Psi_{+}$, the eigenvalue associated with the $z$-depedent profiles, $\lambda^{2}_{+j}$, is always more than zero, thus all these modes are massive. Also, $\lambda^{2}_{+j}=\lambda^{2}_{-j}$ for $j=1,2,..., \lfloor{}\tilde{h}_{\eta_2}\rfloor{}$. If one looks at Eqs.\ \eqref{eq:fermionschrodingeqpmlefty} and \ref{eq:fermionschrodingeqpmrighty}, one sees that the $y$-dependent profiles $f^{m}_{\pm{}L}$ of the chiral modes of $\Psi_{-}$ satisfy the same equation as those of equal mass and opposite chirality in $\Psi_{+}$,  In other words, $f^{m}_{\pm{}L}=f^{m}_{\mp{}R}$, which in turn implies that $\lambda^{1}_{+L,R}=\lambda^{1}_{-R,L}$. Thus the masses for these modes are simply
\begin{equation}
\begin{aligned}
\label{eq:minusleftkkmasses}
m^2_{+Lij} &= \tilde{\lambda}^1_{+Li}k^2+\tilde{\lambda}^2_{+j}l^2, \\
           &= 2ih_{\eta_1}v_{1}k-2i^2k^2+2jh_{\eta_2}v_{2}l-2j^2l^2, \\ 
           &i=1,..., \lfloor{}\tilde{h}_{\eta_1}\rfloor{}, \quad{}j=1,..., \lfloor{}\tilde{h}_{\eta_2}\rfloor{},
\end{aligned}
\end{equation}
and
\begin{equation}
\begin{aligned}
\label{eq:minusleftkkmasses}
m^2_{+Rij} &= \tilde{\lambda}^1_{+Ri}k^2+\tilde{\lambda}^2_{+j}l^2, \\
           &= 2ih_{\eta_1}v_{1}k-2i^2k^2+2jh_{\eta_2}v_{2}l-2j^2l^2, \\ 
           &i=0,1,..., \lfloor{}\tilde{h}_{\eta_1}\rfloor{}, \quad{}j=1,..., \lfloor{}\tilde{h}_{\eta_2}\rfloor{}.
\end{aligned}
\end{equation}

Obviously, these modes should satisfy a 3+1D Dirac equation. The above KG equations and the resultant equations for the profiles give clues for what the form of these should be. Each mode belonging to $\Psi_{-}$ or $\Psi_{+}$ has a $z$-dependent profile $g^{m}_{-}$ or $g^{m}_{+}$ respectively. In turn, each chiral mode inside $\Psi_{-}$ and $\Psi_{+}$ has a $y$-dependent profile $f^{m}_{\pm{}L,R}$, and as we noted above, $f^{m}_{\pm{}L}=f^{m}_{\mp{}R}$. In using this expansion and performing dimensional reduction, one expects a left-chiral mode of $\Psi_{-}$ with a particular mass to not only attain a mass term with the corresponding right-chiral modes of $\Psi_{-}$ but also those of $\Psi_{+}$. This is particularly important when considering that for a left-chiral mode in $\Psi_{-}$ which has $\lambda^1_{-L}=0$ but $\lambda^2_{-}\not=0$ (ie. the mode is a zero mode with respect to the $y$-dependent wall) given that there is no corresponding right-chiral mode of the same eigenvalues and thus mass in $\Psi_{-}$. 
However, there is in $\Psi_{+}$ and thus with such modes there is a single Dirac fermion of the given mass formed from the left-chiral mode in $\Psi_{-}$ and the right-chiral mode of $\Psi_{+}$. Hence, the correct ansatz is that the effective 3+1D mass Lagrangian should be of the form 
\begin{equation}
\label{eq:fermionansatzmasslagrangian}
\mathcal{L} = \begin{pmatrix} \overline{\varphi^{m}_{-L}} & \overline{\varphi^{m}_{+L}} \end{pmatrix} 
              \begin{pmatrix} -\sqrt{\lambda^{1}_{-L}} & \sqrt{\lambda^{2}_{-}} \\
                              \sqrt{\lambda^{2}_{-}}  & \sqrt{\lambda^{1}_{-L}} \end{pmatrix} 
              \begin{pmatrix} \varphi^{m}_{-R} \\
                              \varphi^{m}_{+R} \end{pmatrix}+h.c.
\end{equation}

 For the massive modes with both $\lambda^{1}_{-L}$ and $\lambda^{2}_{-}$ non-zero, one can deduce from the mass matrix in Eq.\ \ref{eq:fermionansatzmasslagrangian} that there exist two Dirac fermion modes of mass $\sqrt{\lambda^{1}_{-L}+\lambda^{2}_{-}}$. After putting this ansatz into the 5D Dirac equation and doing some algebra, one can show that the equations yielded for the $y$ and $z$-dependent profiles are exactly the same as those derived above from the Klein-Gordon equation. 

 In this section we have shown that there exists a single chiral zero mode localized to the domain-wall intersection when the 5+1D fermionic field $\Psi$ is subject to the Yukawa interactions in Eq.\ \ref{eq:fermionlocalizationpotential}. In addition to this single chiral zero mode, there is a single tower of Dirac modes which are zero modes with respect to one wall but not the other, and then for each given squared mass value for which the eigenvalues associated with each wall are both non-zero there exist two Dirac modes. There will also be modes with one of $\lambda^{1}_{\pm{}L,R}$ or $\lambda^{2}_{\pm}$ being more than the maximum value for the localized KK modes and the other corresponding to a value associated with a localized mode; these modes can propogate along one wall and behave as 5D delocalized modes. Modes for which both the eigenvalues $\lambda^{1}_{\pm{}L,R}$ and $\lambda^{2}_{\pm}$ are more than the maximum values for the localized modes are completely delocalized from both walls and can propogate through the entire 6D bulk.

\section{Scalar Localization}
\label{sec:scalarlocalization}

 Scalar localization can be similarly achieved via quartic coupling to the background scalar fields. Generally, in a model that is to be physically viable we are interested in localizing Higgs fields which have gauge charges. Thus for this section we will assume that our candidate scalar field is a complex scalar field $\Phi$. The scalar potential for $\Phi$ is then
\begin{equation}
\begin{aligned}
\label{eq:scalarlocpotential}
V_{\phi} &= \frac{1}{2}\mu^2_{\Phi}\Phi^{\dagger}\Phi+\frac{1}{4}\lambda_{\Phi}(\Phi^{\dagger}\Phi)^2+\frac{1}{2}\lambda_{\Phi\eta_1}\eta^2_1\Phi^{\dagger}\Phi \\
         &+\frac{1}{2}\lambda_{\Phi\chi_1}\chi^2_1\Phi^{\dagger}\Phi+\frac{1}{2}\lambda_{\Phi\eta_1\chi_1}\eta_{1}\chi_{1}\Phi^{\dagger}\Phi+\frac{1}{2}\lambda_{\Phi\eta_2}\eta^2_2\Phi^{\dagger}\Phi \\
         &+\frac{1}{2}\lambda_{\Phi\chi_2}\chi^2_2\Phi^{\dagger}\Phi+\frac{1}{2}\lambda_{\Phi\eta_2\chi_2}\eta_{2}\chi_{2}\Phi^{\dagger}\Phi.
\end{aligned}
\end{equation}
 Assuming that either $\Phi$ has a vanishing vacuum expectation value (VEV) or one of much smaller magnitude than those that the background fields attain (as it would be in the case of an electroweak Higgs boson), we can ignore the quartic self-coupling for $\Phi$ in the determination of the profiles when we do a mode expansion. Hence, we can focus solely on the couplings of $\Phi$ to the background fields and the mass term, and to calculate the profiles we must solve the 5+1D KG equation
\begin{equation}
\begin{aligned}
\label{eq:scalarkglocequation}
&\big[\Box{}+\mu^2_{\Phi}+\lambda_{\Phi\eta_1}\eta_1^2+\lambda_{\Phi\chi_1}\chi^2_1+\lambda_{\Phi\eta_1\chi_1}\eta_{1}\chi_{1} \\
&+\lambda_{\Phi\eta_2}\eta^2_2+\lambda_{\Phi\chi_2}\chi^2_2+\lambda_{\Phi\eta_2\chi_2}\eta_{2}\chi_{2}\big]\Phi=0.
\end{aligned}
\end{equation}
 Again assuming the same perpendicular solution for the background that we assumed in the previous section, we expand $\Phi$ as a series of modes
\begin{equation}
\label{eq:scalarmodeexpansion}
\Phi{}(x_{\mu}, y, z)=\sum_{\substack{m}}p_{m}(y)q_{m}(z)\phi_{m}(x_{\mu}),
\end{equation}
where the $\phi_{m}(x_{\mu})$ are 3+1D scalar modes satisfying the KG equation $\Box{}\phi_{m}(x_{\mu})=-m^2\phi_{m}(x_{\mu})$and the $p_{m}(y)$ and $q_{m}(z)$ are the associated profiles along the $y$ and $z$ directions respectively. Subsitituting this expansion into Eq.\ \ref{eq:scalarkglocequation}, and then demanding that profiles satisfy the Schr$\ddot{o}$dinger equations
\begin{subequations}
\label{eq:scalarprofileschrodinger}
\begin{align}
&\big[-\frac{d^2}{dy^2}+\lambda_{\Phi\eta_1}\eta^2_{1}(y)+\lambda_{\Phi\chi_1}\chi^2_{1}(y) \nonumber \\
&+\lambda_{\Phi\eta_1\chi_1}\eta_{1}(y)\chi_{1}(y)\big]p_{m}(y)=\lambda^1_{m}p_{m}(y) \\
&\big[-\frac{d^2}{dz^2}+\lambda_{\Phi\eta_2}\eta_2^{2}(z)+\lambda_{\Phi\chi_2}\chi^2_{2}(z) \nonumber \\
&+\lambda_{\Phi\eta_2\chi_2}\eta_{2}(z)\chi_{2}(z)\big]q_{m}(z)=\lambda^2_{m}q_{m}(z), 
\end{align}
\end{subequations}
reduces Eq.\ \ref{eq:scalarkglocequation} to a relation between the masses of the KK modes $m$ to the eigenvalues $\lambda^1_{m}$, $\lambda^2_{m}$ and the 5D bare mass $\mu_{\Phi}$
\begin{equation}
\label{eq:scalarkkmasses}
m^2=\mu^2_{\Phi}+\lambda^1_{m}+\lambda^2_{m}.
\end{equation}
 Working in the non-dimensionalised coordinates $\tilde{y}_1=\tilde{y}=ky$, $\tilde{y}_2=\tilde{z}=lz$, using the notation $p^1_m=p_m$ and $p^2_m=q_m$, and given the perpendicular solution given in Eq.\ \ref{eq:perpendicularsolution}, the Schr$\ddot{o}$dinger equations \ref{eq:scalarprofileschrodinger} can both be rewritten in the form
\begin{equation}
\begin{gathered}
\label{eq:hyperbolicscarfpotential}
-\frac{d^2p^i_m}{d\tilde{y}_i^2}+V^i_{HS}(\tilde{y_i})p^i_m(\tilde{y}_i)= E^{i}_{m}p^i_m(\tilde{y}_i), \\
V^i_{HS}(\tilde{y}_i) = a^{2}_{i}+\big(b^{2}_{i}-a^{2}_{i}-a_{i}\big)\sech^2{(\tilde{y}_i)} \\
                  +b_{i}(2a_{i}+1)\sech{(\tilde{y}_i)}\tanh{(\tilde{y}_i)},
\end{gathered}
\end{equation}
where the $a_i$ and $b_i$ are defined as
\begin{equation}
\begin{aligned}
\label{eq:scalarhyperbolicscarfparameters}
a_i &= \frac{1}{2}\bigg(-1+\big(2[(\tilde{\lambda}_{\Phi\chi_i}-\tilde{\lambda}_{\Phi\eta_i}-\frac{1}{4})^2+\tilde{\lambda}^2_{\Phi\eta_i\chi_i}]^{\frac{1}{2}} \\
    &-2\tilde{\lambda}_{\Phi\chi_i}+2\tilde{\lambda}_{\Phi\eta_i}+\frac{1}{2}\big)^{\frac{1}{2}}\bigg),                                                         \\
b_i &= \frac{\tilde{\lambda}_{\Phi\eta_i\chi_i}}{2a_{i}+1},
\end{aligned}
\end{equation}
the nondimensionalized versions of the original eigenvalues and quartic scalar couplings are defined as
\begin{equation}
\begin{gathered}
\label{eq:dimensionlessquarticcouplings}
\tilde{\lambda}^{1}_{m} = \frac{\lambda^{1}_{m}}{k^2}, \quad{} \tilde{\lambda}_{\Phi{}\eta_{1}} = \frac{\lambda_{\Phi{}\eta_{1}}v_{1}^2}{k^{2}}, \quad{} \tilde{\lambda}_{\Phi{}\chi_{1}} = \frac{\lambda_{\Phi{}\chi_{1}}A_{1}^2}{k^{2}}, \\
\tilde{\lambda}_{\Phi{}\eta_{1}\chi_{1}} = \frac{\lambda_{\Phi{}\eta_{1}\chi_{1}}v_{1}A_{1}}{k^{2}}, \quad{} \tilde{\lambda}^{2}_{m} = \frac{\lambda^{2}_{m}}{l^2}, \quad{} \tilde{\lambda}_{\Phi{}\eta_{2}} = \frac{\lambda_{\Phi{}\eta_{2}}v_{2}^2}{l^{2}},  \\
\tilde{\lambda}_{\Phi{}\chi_{2}} = \frac{\lambda_{\Phi{}\chi_{2}}A_{2}^2}{l^{2}}, \quad{} \tilde{\lambda}_{\Phi{}\eta_{2}\chi_{2}} = \frac{\lambda_{\Phi{}\eta_{2}\chi_{2}}v_{2}A_{2}}{l^{2}},
\end{gathered}
\end{equation}
and the eigenvalues of these potentials $E^{i}_{m}$ given in terms of $\tilde{\lambda}^i_m$, $\tilde{\lambda}_{\Phi\eta_i}$ and $a_{i}$ are
\begin{equation}
\label{eq:hyperbolicscarfevalues}
E^i_{m} = \tilde{\lambda}^i_m-\tilde{\lambda}_{\Phi\eta_i}+a^2_i.
\end{equation}

 Assuming $a_1$ and $a_2$ are positive\footnote{If one of $a_1$ or $a_2$ is negative, then there are no modes localized to the intersection and there will exist modes localized to one wall but delocalized from the other. If both are negative, then all modes are delocalized from both walls.}, these hyperbolic scarf potentials yield a discrete set of modes localized to the intersection of the domain walls. A localized $\phi_m$ mode must clearly have both of $p_m$ and $q_m$ decay to zero as $y\rightarrow\pm\infty$ and $z\rightarrow\pm\infty$ respectively. Since the respective hyperbolic scarf potentials for the $p_m$ and $q_m$ yield $\lceil{}a_1\rceil{}$ and $\lceil{}a_2\rceil{}$ localized functions respectively, there are $\lceil{}a_1\rceil{}\lceil{}a_2\rceil{}$ modes localized to the domain-wall intersection. For each potential, the eigenvalues for the localized modes are known to be
\begin{equation}
\label{eq:hyperbolicscarflocalisedevalues}
E^i_n = 2na_i-n^2,
\end{equation}
for $n=0, 1, ..., \lfloor{}a_i\rfloor{}$, and using Eqs.\ \ref{eq:scalarkkmasses}, \ref{eq:hyperbolicscarfevalues}, and \ref{eq:hyperbolicscarflocalisedevalues} we thus find that the squared masses of the localized 3+1D modes are
\begin{equation}
\label{eq:scalarlocalisedmodemasses}
m^2_{ij} = \mu^2_{\Phi}+\tilde{\lambda}_{\Phi\eta_1}k^2+\tilde{\lambda}_{\Phi\eta_2}l^2-(a_1-i)^2k^2-(a_2-j)^2l^2,
\end{equation}
for $i=0, 1, ..., \lfloor{}a_1\rfloor{}$ and $j=0, 1, ..., \lfloor{}a_2\rfloor{}$.

 There will also exist modes with sufficient average momenta transverse to one domain wall but not the other such that these modes are localized to one wall but not the other. These modes have a profile with one of the quantum numbers described above with respect to one wall but will have energies above that of the most energetic localized mode of the other. These modes are essentially 5D particles. 

 For modes with energies above all localized modes, their profiles along both directions are delocalized and they are thus 6D particles with can propogate along the full extent of the bulk. 

 In any potentially phenomenological model based on this type, we are usually interested in the lowest energy localized mode, as this 4D scalar mode would correspond to Higgs particles in the effective field theory on the domain-wall intersection. In our model, this mode is the $i=0$, $j=0$ mode of Eq.\ \ref{eq:scalarlocalisedmodemasses}. The resultant (non-dimensionalised) profiles for this mode $\tilde{p}_{0}(\tilde{y})$ and $\tilde{q}_{0}(\tilde{z})$ are simply given by
\begin{equation}
\begin{aligned}
\label{eq:higgsyzprofiles}
\tilde{p}_{0}(\tilde{y}) &= \tilde{C}_{0}e^{-a_{1}\log{[\cosh{(\tilde{y})}]}-2b_{1}\arctan{[\tanh{(\tilde{y}/2)}]}},  \\
\tilde{q}_{0}(\tilde{z}) &= \tilde{D}_{0}e^{-a_{2}\log{[\cosh{(\tilde{z})}]}-2b_{2}\arctan{[\tanh{(\tilde{z}/2)}]}}.
\end{aligned}
\end{equation}
Here, $\tilde{p}_{0}(\tilde{y})=k^{-\frac{1}{2}}p_{0}(y)$ and $\tilde{q}_{0}(\tilde{z})=l^{-\frac{1}{2}}q_{0}(z)$.

\section{Conclusion}
\label{sec:conclusion}

 In this paper, we have generated a rare analytic solution to a $\mathbb{Z}_{2}\times{}\mathbb{Z}_{2}$-invariant scalar field theory with four real scalar fields in 5+1D spacetime describing a pair of intersecting domain walls with internal structure. We found that with respect to the desirable perpendicular solution, there also existed a class of solutions describing kink-lump solutions which intersect at an angle between 0 and 90 degrees as well as a solution where the walls are parallel and that these solutions were energy degenerate. We then argued that there exists a conserved topological charge related to the one-dimensional boundary of the $y-z$ plane which differed between the intersecting solutions and the parallel solution, meaning that the intersecting solutions cannot evolve to the parallel one despite this energy degeneracy. We also gave an argument as to why the perpendicular solution might be energetically favorable to the solutions with intersection angle less than 90 degrees in a nearby 
region of parameter space.

 In addition to finding this solution, we showed in the case that the two domain walls were perpendicular that fermions and scalars could be dynamically localized to the intersection of the two walls. We found that coupling a 6D fermionic field to one kink-lump pair with ordinary scalar Yukawa couplings and to the other with pseudoscalar Yukawa couplings allowed the fermionic sector to be invariant under the full $\mathbb{Z}_{2}\times{}\mathbb{Z}_{2}$ symmetry and resulted in the localization of a 4D chiral zero mode on the intersection of the kinks, followed by a tower of localized KK modes and 5D and 6D delocalized modes. 

 Standard quartic couplings of a complex scalar field to the background scalar fields resulted in a tower of localized 4D localized scalar modes with the squared masses starting from some potentially non-zero value. This result is similar to the result for scalar localization to a single domain wall in 5D. Furthermore, the squared mass of the lowest energy scalar mode can be negative, allowing the possibility of the localized scalar field inducing spontaneous symmetry breaking should that scalar field transform under a non-trivial gauge group representation. 

 Localization of gravity and gauge bosons is left to later work. Localization of gravity would involve searching for a similar solution to the 6D Einstein field equations as well as the Einstein-Klein-Gordon equations. These equations are highly non-linear and difficult to solve so it remains to be seen if an analytic solution could be found. It could also be the case that such a solution could have qualitative differences as in principle the two domain-wall branes should interact gravitationally, whereas in this flat space case the net interaction between the two kink-lump solutions was zero. 

 For the localization of gauge bosons, we conjecture that the Dvali-Shifman mechanism works in 5+1D spacetime. This ultimately depends on whether or not non-Abelian gauge theories are confining in 5+1D. In previous work in 4+1D, particularly in the $SU(5)$ model \cite{firstpaper}, the Dvali-Shifman mechanism was facilitated by the addition of a scalar field transforming under the gauge group which attained a lump-like profile and induced symmetry breaking in the interior of the domain wall. Hence, the background solution of this model was of the kink-lump type discussed in this paper, and the additional fields $\chi_{1}$ and $\chi_{2}$ which attained lump-like profiles in the 6D model are potentially well motivated. With the assignment of gauge representations, these fields can break a gauge group to different subgroups which then \emph{clash} at the intersection, leading to further symmetry breaking and a different realization of the Clash-of-Symmetries mechanism \cite{clashofsymmetries, e6domainwallpaper}. 
We will discuss in detail this realization of the Clash-of-Symmetries mechanism and its applications to model building in a later paper.

\begin{acknowledgments}
This work was supported in part by the Australian Research Council. We thank Damien George for some useful discussions.
\end{acknowledgments}

\newpage{}

\bibliographystyle{ieeetr}
\bibliography{bibliography2.bib}

\end{document}